 \documentclass[10.5pt]{article}
\usepackage{amsmath}
\usepackage{bm}
\usepackage[utf8]{inputenc}
\usepackage{graphicx}
\usepackage{geometry}
\usepackage{enumerate}
\usepackage{amssymb}
\usepackage{subfig}
\usepackage{float}
\usepackage[section]{placeins}
\usepackage{tabto}
\usepackage{multirow}
\usepackage[svgnames]{xcolor}
\usepackage[pagestyles]{titlesec}
\usepackage{lipsum}
\usepackage[acronym]{glossaries}
\usepackage{url}
\usepackage{hyperref}
\usepackage{amsmath}
\usepackage[numbers,sort&compress]{natbib}
\usepackage{comment}
\newcommand{\bea}{\begin{eqnarray*}}    
\newcommand{\eea}{\end{eqnarray*}}
\newcommand{\be}{\begin{eqnarray}}    
\newcommand{\ee}{\end{eqnarray}}
\usepackage{amsthm}
\newtheorem{lemma}{Lemma}
\newtheorem{theorem}{Theorem}
\newtheorem*{definition}{Definition}
\newtheorem{corollary}{Corollary}

\usepackage{bbm}
\graphicspath{{Images/}}

\usepackage{natbib}
\begin{document}

\title{An integrated approach to test for missing not at random}
\author{ J. NOONAN$^{1}$ \and A. A. ADEDIRAN$^2$ \and R. MITRA$^{3}$ \and S. BIEDERMANN$^4$}
\date{%
    $^1$  {Noonanj1@cardiff.ac.uk}, School of Mathematics, Cardiff University, U.K\\%
     $^2${A.A.adediran@soton.ac.uk}, School of Mathematical Sciences, University of Southampton, U.K.\\ 
      $^3$ {Robin.mitra@ucl.ac.uk},  Department of Statistical Science, UCL, U.K.\\  %
    $^4${Stefanie.biedermann@open.ac.uk}, School of Mathematics and Statistics, The Open University, U.K.\\[2ex]
}
\maketitle

\begin{abstract}
\noindent Missing data can lead to inefficiencies and biases in analyses, in particular when data are missing not at random (MNAR). It is thus vital to understand and correctly identify the missing data mechanism. Recovering missing values through a follow up sample allows researchers to conduct hypothesis tests for MNAR, which are not possible when using only the original incomplete data. Investigating how properties of these tests are affected by the follow up sample design is little explored in the literature. Our results provide comprehensive insight into the properties of one such test, based on the commonly used selection model framework. We determine conditions for recovery samples that allow the test to be applied appropriately and effectively, i.e. with known Type I error rates and optimized with respect to power. We thus provide an integrated framework for testing for the presence of MNAR and designing follow up samples in an efficient cost-effective way.
The performance of our methodology is evaluated through simulation studies as well as on a real data sample.
\end{abstract}

\vfill

\section{Introduction}

When analysing data, missing values are often encountered. 
Various methods to handle missing data have been proposed in the literature \citep{little2002bayes}, and determining the appropriate method to use depends on the specific nature of the missing data. In particular, missing data mechanisms \citep{rubin1976inference}, which model the process generating the missing values, are fundamental in formulating appropriate analysis methods for the incomplete data. While it is typical to assume the mechanism is missing at random (MAR), the reasons for this are often due to necessity rather than a firm belief this assumption holds. The alternative, which is to assume missing not at random (MNAR) poses significant challenges. Its presence introduces biases in the analysis, and is an untestable assumption based on the original incomplete sample \citep{little2002bayes}. As effects of MNAR should only be corrected if we have good confidence in its presence and functional form, it is important to first design an appropriate test for detecting MNAR. However, the literature on tests for MNAR is very sparse and not well developed, which motivate the findings presented in this article.

While most methods address missing values post hoc, this is typically due to the inability of implementing (or failure to plan) for this eventuality. A well conducted scientific study, and indeed common sense, suggests accounting for missing values prior to data collection is good practice, as it allows the study design to be planned accordingly. For example, the few papers that develop theory to construct optimal study designs accounting for the potential of non-response 
demonstrate the benefits possible \citep{imhof2002,lee2018a,lee2019}. 

In a similar vein, planning a follow up sample to recover some of the missing values is the most effective way to learn about the existence of MNAR, and avoids having to rely on a number of (unverifiable) assumptions about the missing data mechanism. The value of follow up sampling (often denoted as double sampling, call backs, or repeated attempt design) to address problems caused by missing data, has been recognized in various fields, such as survey sampling \citep{elliott2000,guan2018,miao2021,alho1990,drew1980,qin2014}, survival studies \citep{frangakis2001,an2009,qian2020}, and randomized experiments \citep{jackson2010,aronow2015,daniels2015,coppock2017}. However, much of the literature focuses on using follow up samples to facilitate the applied analysis of interest rather than seeking to learn about the missing data mechanism. When follow up has been used to infer properties of the mechanism, little attention has been given to determining the effect different follow up strategies have on inferences. Typically, a simple follow up, e.g. sampling uniformly from the non-responders or, in the case of repeated attempt design, sampling all non-responders, is implicitly assumed, and in many instances the recovery scheme is not explicitly considered. While this may be appropriate in some scenarios, in many settings, the amount of recovery is likely to be limited due to practical or ethical constraints, leading to modest recovery samples. In these cases, a random follow up may be spread too thinly over the space of missing records to allow a statistical test to detect the presence of MNAR with any degree of certainty. However, a {\em carefully selected} follow up sample, may result in substantial efficiency gains permitting a test for MNAR with higher power. 

There has been little to no attention investigating how follow up sampling can be used to minimize the fundamental uncertainty concerning the presence of MNAR in incomplete data. \cite{levis2022} recognize uncertainty surrounding MNAR in the context of estimating treatment effects in observational studies with follow up, but address this using a data-adaptive estimator, without optimizing tests for MNAR and associated theory in this paradigm. 
In this paper, we  propose a fully integrated approach to test for MNAR that involves constructing follow up sample designs to optimize the ability of a statistical test to detect the presence or absence of MNAR. Our framework is underpinned by results that rigorously establish the proposed test's theoretical properties, allowing it to be implemented with confidence. 


Specifically, after conducting a follow up sample, we assume a regression model is formulated to test for MNAR based on the originally observed sample together with the newly recovered values \citep{carpenter2012multiple}. The choice of regression model to use depends on whether the Pattern Mixture or Selection Model Framework (SMF) is adopted. In this article we choose the latter as it more closely corresponds with \cite{rubin1976inference}'s definition of mechanisms and allows intuitive interpretation of the results. Here, a binary response regression can test for MNAR (typically through a logit link) where the outcome (missing indicator) is modeled as a function of the covariates \textit{and the response variable} (which now comprises a mix of observed and recovered values). The test is formally defined in Section 2 but, intuitively, detecting MNAR is equivalent to testing whether the regression coefficients on terms involving the response variable are 0 or not. 

The natural question of which missing records to follow up then arises, particularly in the practical setting where only a modest follow up is possible. The choice of follow up sample could affect the power of this test, but also, crucially, the possibly non-representative sample of observed plus recovered values could alter the distributional properties of the data and thus affect estimation of model parameters, leading to potential bias in the test's Type I error. This phenomenon has been noted in other settings involving binary response models fit to subsamples \citep{wang2018optimal,wang2019more,yao2021review}. In these papers, to compute the optimal weights used within the subsampling procedure, all values must be observed. Therefore an immediate application of the results in \cite{wang2018optimal}, \cite{wang2019more} and \cite{yao2021review} is not clear in the presence of missing data.

Our research addresses the above open challenges in the following ways.
Firstly, utilizing the SMF to test for MNAR with any arbitrary link function that maps from $[0,1]$ to $\mathbb{R}$, we derive conditions the follow up design must satisfy to ensure validity of inferences, such as that it achieves desired Type I error rates. Secondly by treating the power of the test as a function of the recovery sample design, we develop a framework that allows the recovery sample to maximize the power of the test. Thirdly we show that the logit link function has some appealing properties over other link functions that suggest its use, in the absence of information recommending a different link function. We thus comprehensively address the various different facets associated with this gap in the literature on tests for MNAR. 






{The remainder of this article is organized as follows. In Section 2, we formulate the problem and introduce the test for MNAR based on the SMF. Section 3 contains our main theoretical results. In particular, we derive expressions for the distributions of the augmented data in terms of mixture distributions and employ these to construct a formula for the missing data mechanism in the augmented data, which provides vital insight into the estimation problem of the MNAR model and the SMF test. In Section 4, we propose an optimality criterion and provide an algorithm to construct cost efficient recovery designs. In Section 5, our methodology is illustrated through two examples and a robustness study of our designs is presented. The practical relevance of our approach is demonstrated through a real data example in Section 6, followed by concluding remarks in Section 7. The proofs of our results can be found in the Appendix.}

\section{Testing for MNAR}
\subsection{Problem formulation}	 \label{sec:problem_formulation}
To motivate our methodology, consider a setting involving a univariate response, $Y$, and a $p$-dimensional covariate vector ${\bm{X}=(X_1,\ldots, X_p)^T}$, where
\be\label{Main_regression_model}
Y \vert (\bm{X}=\bm{x}) &\sim& N(\beta_0+\bm{\beta}^T\bm{x},\sigma^2_y) \,,
\ee
with $\bm{x} = (x_1,\ldots,x_p)^T$, $\beta_0\in\mathbb{R}$ and  $\bm{\beta} = (\beta_1\ldots, \beta_p)^T \in \mathbb{R}^{p}$. Let $\bm{X}$ be a random vector with any arbitrary joint density function $f(\bm{x})$. 
For the construction of designs in Section~\ref{sec:design}, we will assume $\bm{\beta},\beta_0$ and $\sigma^2_y$ are known. Whilst this {assumption may be viewed as restrictive, it adheres to the well established principle of locally optimal designs \citep{chernoff1953}, and through simulations provided in Section~\ref{sec:robustness} we establish that our designs are quite robust to misspecification of these values. Moreover, locally optimal designs play an important role as benchmarks for design assessment.} 
We assume realizations of $\bm{X}$ are always fully observed and that missing values only occur within the response $Y.$ Let $M$ denote an indicator random variable equal to one if $Y$ is missing and zero if observed. The conditional distribution of $M$, ${\rm Pr} (M=1 \vert \bm{X}= \bm{x}, Y = y)$, determines the missing data mechanism \citep{rubin1976inference}. Under MAR, we have $
{\rm Pr} (M=1 \vert \bm{X}=\bm{x}, Y = y) = {\rm Pr} (M=1 \vert \bm{X}=\bm{x}) \,;
$
i.e. $M$ is conditionally independent of $Y$ given the covariates. Under MNAR, this property does not hold. In this paper, we assume the missing mechanism has the form
\be\label{SMF_logit}
g( {\rm Pr}{(M = 1 \vert \bm{X}=\bm{x},Y=y)} )= \bm{w}^T\bm{\lambda} + \bm{z}^T\bm{\psi}\,.
\ee
where $g$ is any arbitrary link function for a generalized linear model (GLM) with a binary response, $\bm{w}$ is a $q$-dimensional vector whose components could depend on functions of $\bm{x}$ but not $y$, and $\bm{z}$ is an $s$-dimensional vector whose components additionally depend on 
functions of $y$, for example an interaction $x_{i}y$, and/or a function of $y$ on its own. Without loss of generality, we will assume the first component of $\bm{w}$ is equal to one; this corresponds to assuming an intercept term is present in the missing mechanism.  Here $\bm{\lambda}$ and $\bm{\psi}$ are the unknown vectors of coefficients. 
Typical choices of $g$ include the logit, probit, and complementary log-log link functions. By taking the inverse of the link function, an equivalent form of \eqref{SMF_logit} that models the conditional distribution directly is
\be\label{SMF_logit2}
{\rm Pr}{(M = 1 \vert \bm{X}=\bm{x},Y=y)} = g^{-1}(\bm{w}^T\bm{\lambda} + \bm{z}^T\bm{\psi})\,.
\ee
For the common choices of link functions, their inverses are given by
\bea
g^{-1}(t) =  
  \left\{\begin{array}{lr}
        {\exp(t)}/({1+\exp({t}})) = \mbox{expit}(t), & \text{for the logit model }  \\
        \Phi(t), & \text{for the probit model }  \\
        1-\exp({-\exp(t)}), &  \text{for the complementary log-log model, }
       \end{array} \right.
\eea
where $\Phi(t)$ is the standard normal distribution function.

{Determining the presence and type of MNAR thus involves determining the value of $\bm{\psi}$, with MAR occurring only when $\bm{\psi} = \bm{0}$. However, this parameter is inestimable based on the original sample as $Y$ is only observed for $M = 0$.}
In order to address this we formulate a two-stage experiment to collect the data. At stage one, let $Y_1,\ldots,Y_n$, $\bm{X}_1,\ldots,\bm{X}_n$ and $M_1,\ldots,M_n$ denote samples of size $n$ from the model in \eqref{Main_regression_model}.  Within the sample of $Y_1,\ldots,Y_n$, suppose $n_{obs} < n$ of $Y_i$ are observed, meaning $n_{miss} := n-n_{obs}$ observations of $Y_i$ are missing. Without loss of generality, we assume the first $n_{obs}$ of $Y_1,\ldots,Y_n$ are observed, 
meaning $Y_{n_{obs}+1},\ldots,Y_{n}$ are all initially missing.
At stage 2, assume resources permit follow up of a number of experimental units with missing responses to obtain (recover) their responses, e.g. through home visits to patients in a clinical trial or follow-up telephone calls in a survey. We denote the number of recovered responses by $n^*$, where $n^*\leq n_{miss} $. As the number of recovered responses is a proportion of the missing observations, we can express $n^*= \lceil c_1\cdot n_{miss} \rceil$ with $0<c_1\leq 1$. We assume the choice of which responses to recover is in the practitioner's control, giving rise to the concept of a recovery design.
\begin{definition}
A recovery design ${\bf {R}}= {\bf {R}}(n^*)$ is a subset of size $n^*$ from ${{\bf M}:=\{ {n_{{obs}+1}},n_{{obs}+2} \ldots, n \}}$.
\end{definition}
The follow-up design ${\bf {R}}$ instructs the practitioner what missing values to recover as follows. Specifically, a set of $n^*$ observations among the $n_{miss}$ missing observations  to be recovered. To facilitate notation, assume these units are relabelled so that the recovery design ${\bf {R}}=\{k_1,k_2,\ldots,k_{n^*} \}$. This results in the recovery of the previously missing response vector ${\bf Y_{{\bf {R}} }} := ( Y_{k_1}, \ldots,Y_{k_{n^*}}) $. By combining this with the observed responses ${\bf Y_{O}} := ( Y_{1},Y_{2},\ldots, Y_{{n}_{obs}}  )$, we obtain the augmented responses $ {\bf Y_{A}:=(\bf Y_{O}}, {\bf Y_{{\bf R}}})^T$. 
The analogous matrix and vector of augmented data for the covariates and the indicator random variables are, respectively: $ {\bf X_{A}}:=[{\bf X_{O}}, {\bf X_{R}}]^T$ with ${\bf X_{R} }:= [ \bm{X}_{k_1}, \ldots,\bm{X}_{k_{n^*}}]$ and ${\bf X_{O}} := [ \bm{X}_{1},\ldots, \bm{X}_{{n}_{obs}}  ]$; ${\bf M_{A}}:=({\bf M_{O}}, {\bf M_{R}})^T$ with ${\bf M_{O}}=(0,\ldots,0)$ and ${\bf M_{R}}=(1,\ldots,1)$. 

With the above formulation we are now able to make inferences about the missing data mechanism. In particular, we can address the following two objectives: 1) Obtain unbiased estimates of parameters characterizing the missing data mechanism with a view to testing for MNAR, and 2) Improve the power of a test to detect MNAR if present. 

A key complication 
is that, even if the missing data mechanism can be well modeled by \eqref{SMF_logit} based on the complete sample, the observed plus recovered sample may not be representative of the wider population. Thus the original model for the missing data mechanism may no longer be plausible, both in terms of its functional form as well as its parameter values.



Assuming a recovery design 
takes a random sample within a specified $p$-dimensional region of the covariate space, the key developments in this paper are as follows. Firstly, for a missing mechanism of the form in \eqref{SMF_logit2} with logit link,
we show that the {\em entire sample of} augmented data (i.e. observed plus recovered) restricted to this space maintains the $expit$ functional form, but with a modified value of $\bm{\lambda}$. Secondly, for any other link function, with the same recovery design, 
we determine a {\em randomly sampled pre-specified proportion of the observed data} restricted to this space must be taken in order to preserve the mechanism's functional form. 
From these we can establish necessary properties of the statistical test enabling it to be used with confidence, such as the Type I error rate. Furthermore, the inherent benefit with using a logit link (in the absence of evidence to support an alternative link) is evident. The logit link permits all augmented data within the design region to be used for inference while other link functions may require subsampling of the observed record pool to ensure an appropriate statistical test. Deriving the relevant theory leads us to the third key development, which provides a framework for optimizing the power of the test by considering power as a function of the recovery design.

\subsection{A test for MNAR with logistic regression}\label{sec:test}
For $Y_i^* \in  {\bf Y_{A}}$, let ${\bf X^*_{A}}_{,i}$  be the corresponding $i^{th}$ row in the matrix ${\bf X_{A}}$  and let $M_i^* \in  {\bf M_{A}}$ be the corresponding indicator value. Here, the superscript $^*$ is to emphasise we are considering the augmented data. Let $\bm{w}_i$ and $\bm{z}_i$ be the values of $\bm{w}$ and $\bm{z}$ at observation $i$ of the augmented data.
A test for MNAR utilizing the SMF fits the following model:
 \begin{equation}\label{SMF_test}
			{\rm Pr}{(M_i^* = 1 \vert {\bf X^*_{A}}_{,i}={\bf x},Y_i^*=y)} = g^{-1}(\bm{w}_i^T\bm{\lambda}_A + \bm{z}_i^T\bm{\psi}_A)\,.
\end{equation}
 The parameters $\bm{\lambda}_A$ and $\bm{\psi}_A$ are unknown regression coefficients based on the augmented data. To estimate or perform inference on the original parameters $\bm{\lambda}$ and $\bm{\psi}$, one must determine the relation to the parameters $\bm{\lambda}_A$ and $\bm{\psi}_A$. In Section~\ref{sec:mixture}, we show that if the augmented data is constructed in a certain manner, for the logit link function we have the relation $\bm{\psi}_A=\bm{\psi}$ and $\bm{\lambda}_A=\bm{\lambda}+(constant,0,\ldots,0)^T \in {\mathbb{R}^q}$; recall that without loss of generality we assume the first element of $\bm{w}_i$ is equal to one and corresponds to the intercept term in the GLM.  Determining whether MNAR is present or not is then equivalent to testing the hypothesis $\bm{\psi}_A=0$. For any other link function, we can ensure $\bm{\lambda}$ and $\bm{\psi}$ are maintained at the loss of some information.

To perform such a test, we appeal to the likelihood ratio test. The log-likelihood ratio test statistic for testing a general hypothesis $\bm{{\psi}}=\bm{{\psi_0}} $ is given by $2[ l( \bm{\hat{\psi}}, \bm{\hat{\lambda}}_A )-l( \bm{{\psi_0}}, \bm{\hat{\lambda}_0} ) ]$, where $l$ denotes the log likelihood function based on a sample of size $n_A$ (the number of observations in the augmented data) and  $( \bm{\hat{\psi}}, \bm{\hat{\lambda}}_A )$ and $\bm{\hat{\lambda}_0}$ denotes the maximum likelihood estimators under the alternative and null models, respectively. Under the null hypothesis, we approximate the distribution of the likelihood ratio statistic by a central chi-square distribution with $s$ degrees of freedom, i.e. the classical statistical approximation.

\section{Fundamental theoretical findings}

In this section, we formulate the theoretical foundations of the paper. One of the key messages from this section is that
the augmented data used for determining whether MNAR is present represents a sample from a mixture distribution. This arises from a weighted combination of the distribution of the observed data and the distribution of the missing data. This is formalized in the next section.

\subsection{A mixture distribution for the augmented data}\label{sec:mixture}

Let ${\cal C}_{A}\subseteq \mathbb{R}^p$ be  a $p$-dimensional region constructed by the cartesian product of intervals in $\mathbb{R}$ of positive length. For example, we could have ${\cal C}_{A}=\prod_{j=1}^{p} [a_j,b_j]$ with $a_j<b_j$ and $a_j,b_j \in \mathbb{R}$; in this example ${\cal C}_{A}$ becomes a $p$-dimensional hypercuboid if $a_j$ and $b_j$ are finite. However, we could also permit more general sets such as e.g. ${{\cal C}_{A}=\prod_{j=1}^{p} [a_j,b_j]\cup [e_j,f_j]}$ with $e_j<f_j$ and $e_j,f_j$ $\in \mathbb{R}$. The region ${\cal C}_{A}$ will be used to instruct the recovery design  ${\bf {R}}$. In particular, we will only recover $n^*$ missing responses  $Y_{n_{obs}+1},\ldots,Y_{n}$ whose corresponding $p$-dimensional covariate vectors $\bm{X}_{n_{obs}+1},\ldots,\bm{X}_n$ lie within ${\cal C}_{A}$. For the random vector $\bm{X} =(X_{1},\ldots,X_{p}) \in \mathbb{R}^p$ introduced in Section~\ref{sec:problem_formulation}, define the intersection of events:
$$
{\cal M}_O:= \{M=0\} \cap \{\bm{X} \in {\cal C}_{A} \}  \quad \mbox{and} \quad
{\cal M}_R:=  \{M=1\} \cap\{\bm{X} \in {\cal C}_{A} \}  \,.
$$
We assume ${\cal C}_{A}$ is chosen so that  ${\rm Pr}({\cal M}_O)>0$. We could also assume ${\cal C}_{A}$ is chosen so that
\begin{equation} \label{key_condition}
{\rm Pr}({\cal M}_R)= c_1\cdot{\rm Pr}(M=1).
\end{equation}
These requirements ensure enough observations (missing and observed) are present within the region where the augmented data lie, as well as ensuring the recovered sample comprises a fixed proportion, $c$, of the missing values. 
However, when recovery designs involve random selection (without replacement) amongst the missing values, 
(\ref{key_condition}) can be relaxed to permit ${\cal C}_{A}$ such that 
\begin{equation} \label{key_condition2}
{\rm Pr}({\cal M}_R)\geq c_1\cdot{\rm Pr}(M=1).
\end{equation}

Note, if we set ${\cal C}_{A}=\mathbb{R}^p$ such that ${\rm Pr}({\cal M}_R)={\rm Pr}(M=1)$, then ${\bf {R}}$ would be a random sample of size $n^*$ from all the missing responses. Such a design will be referred to as a random recovery design and will act as a benchmark design.

Out of the observed data whose covariates lie within ${\cal C}_{A}$, we will introduce the capability of bringing a random sample of a certain proportion into the augmented data. More specifically, let $0<c_2\leq 1$ be the proportion of observed data whose covariates lie within ${\cal C}_{A}$ that are used within the augmented data. If $c_2=1$, then all of the observed observations that lie in ${\cal C}_{A}$ will be used. If $c_2=0.5$, then 50\% of observed data lying within ${\cal C}_{A}$ is included in the augmented data, provided they are sampled randomly from all the observed covariates that are contained in ${\cal C}_{A}$. As will become clear shortly, if $c_2$ is chosen carefully, the original missing mechanism can be maintained in the augmented data for any GLM link function. With this framework the following results can be derived.

\begin{lemma}\label{main_lemma} The indicator variable for the augmented data, which we denote by $M_A$, satisfies 
\bea
M_A = \begin{cases}
    1 & \text{ with probability  }\, \dfrac{c_1\cdot{\rm Pr}(M=1)}{c_1\cdot {\rm Pr}(M=1)+c_2\cdot {\rm Pr}({\cal M}_O)} \vspace{0.2cm} \\
    0& \text{ with probability }\, \dfrac{c_2\cdot {\rm Pr}({\cal M}_O )}{c_1\cdot {\rm Pr}(M=1)+c_2\cdot {\rm Pr}({\cal M}_O )}\,.
      \end{cases} 
\eea 

\end{lemma}

The proof of Lemma~\ref{main_lemma} can be found in the Appendix.

Accordingly, define the random variables and random vectors
$$
Y_O :=Y \, \vert  \, {\cal M}_O; \quad 
Y_R := Y \, \vert  \, {\cal M}_R; \quad
\bm{X}_O :=\bm{X} \, \vert \, {\cal M}_O; \quad
\bm{X}_R := \bm{X} \, \vert \, {\cal M}_R \,.
$$

\begin{lemma}\label{mixture_formulation}
The augmented response/covariates are realizations from random variable/vector:
\be\label{key_mixture}
Y_A &:=& (1-M_A)Y_O+M_AY_R\\
\bm{X}_A &:=& (1-M_A)\bm{X}_O+M_A\bm{X}_R \, \label{key_mixture2}\,.
\ee

\end{lemma}

These expressions arise by construction, 
being a combination of data with observed outcomes and data with recovered outcomes. Averaging over the distribution of $M_A$, Lemma~\ref{mixture_formulation} can equivalently express the distributions of $Y_A$ and $\bm{X}_{A}$ as mixtures satisfying:
\bea
\mu_{Y_A}(\cdot) &:=& {\rm Pr}(M_A=0)\cdot \mu_{Y_O}(\cdot)+{\rm Pr}(M_A=1)\cdot \mu_{Y_R}(\cdot)\\
\mu_{\bm{X}_{A}}(\cdot) &:=& {\rm Pr}(M_A=0)\cdot \mu_{\bm{X}_O}(\cdot)+{\rm Pr}(M_A=1)\cdot \mu_{\bm{X}_R}( \cdot) \,,
\eea
where,  $\mu_{Y_A}(\cdot)$, and $\mu_{\bm{X}_{A}}(\cdot)$, are probability measures for the random variable/vector, $Y_A$ and $\bm{X}_{A}$ respectively, and similarly define probability measures $\mu_{Y_O}(\cdot)$, $\mu_{\bm{X}_{O}}(\cdot)$ $\mu_{Y_R}(\cdot)$, $\mu_{\bm{X}_{R}}(\cdot)$.

Using the above, together with the problem formulation, we state the key theorem, followed by two corollaries, below.


\begin{theorem}\label{main_theorem}
For $0<c_1\leq 1$ and $0<c_2\leq 1$, provided ${\cal C}_{A}$ satisfies \eqref{key_condition} or \eqref{key_condition2}, then for ${\bm{x} := (x_1,\ldots,x_p)}$ the missing data mechanism in the augmented data has the following form. For $\bm{x} \in  {\cal C}_{A}$ and any link function $g$:
\begin{eqnarray*}
{\rm Pr}(M_A = 1 \, \vert \, \bm{X}_A = \bm{x},  Y_A = y)= 
\dfrac{c^{*}{\rm Pr}(M=1|\bm{X} = \bm{x},Y=y)}{c^{*}{\rm Pr}(M=1|\bm{X} = \bm{x},Y=y)+ {\rm Pr}(M=0|\bm{X} = \bm{x}, Y=y)} \, ,
\end{eqnarray*}
where $c^{*}= {c_1\cdot {\rm Pr}(M=1)}/({c_2\cdot {\rm Pr}(M=1,\bm{X}\in {\cal C}_{A})})$\,. \ Otherwise zero.
\end{theorem}

The proof is provided in the Appendix. 
An important corollary of this theorem is:.
\begin{corollary}\label{c_2_corr}
Under the conditions of Theorem 1, provided
\bea
c_2={c_1\cdot {\rm Pr}(M=1)}/{ {\rm Pr}(M=1,\bm{X}\in {\cal C}_{A})} \,,
\eea
then for $\bm{x} \in  {\cal C}_{A}$, we have ${\rm Pr}(M_A = 1 \, \vert \, \bm{X}_A = \bm{x},  Y_A = y)= {\rm Pr}(M = 1 \, \vert \, \bm{X} = \bm{x},  Y = y),$ otherwise zero.
\end{corollary}
Given a recovery proportion $c_1$, Corollary~\ref{c_2_corr} provides the value of $c_2$, the proportion of observed data lying in ${\cal C}_{A}$ to randomly sample when constructing the augmented data, in order to maintain the original missing mechanism. If $c_2$ does not satisfy the requirements of Corollary~\ref{c_2_corr}, then $c^*\neq 1$ in Theorem~\ref{main_theorem} and there is no cancellation. Consequently, the true missing mechanism in the augmented data will not be of any well known GLM form and therefore obtaining estimates for the parameters and testing the presence of MNAR becomes more complicated. One could build a custom link function from Theorem 1, however easy implementation and the loss of optimized procedures that are present in statistical software packages for well known link functions will likely make analysis cumbersome. An exception to this is the logit link function which maintains the same form regardless of the choice of $c_2$. This is a consequence of the following corollary.
\begin{corollary}\label{cor1}
If the original missing data mechanism utilizes the logit link function:
\begin{equation*} 
{\rm Pr}(M=1 \, \vert \, Y=y,\bm{X} = \bm{x})={\exp(\bm{w}^T\bm{\lambda} + \bm{z}^T\bm{\psi})}/({1+\exp(\bm{w}^T\bm{\lambda} + \bm{z}^T\bm{\psi})})  \, ,
\end{equation*}
then for all $\bm{x} \in { {\cal C}_{A}}$ and any $c_2$, the missing mechanism in the augmented data has the form
\be 
{\rm Pr}(M_A\!=\!1 \, \vert \, Y_A\!=\!y,\bm{X}_A \!=\! \bm{x})={\exp(\bm{w}^T\bm{\lambda}_A + \bm{z}^T\bm{\psi})}/({1+\exp(\bm{w}^T\bm{\lambda}_A + \bm{z}^T\bm{\psi})})  \, ,\label{expit_model}
\nonumber
\ee 
with $\bm{\lambda}_A=\bm{\lambda}+(\log(c^*),0,\ldots,0)^T \in {\mathbb{R}^q}$.
\end{corollary}

While Corollary~\ref{cor1} assumes a MNAR mechanism it also holds under MAR as $\bm{\psi}$ can equal 0. 
The interpretation of Corollary~\ref{cor1} is as follows. If the recovery design is a random sample of the missing responses  $Y_{n_{obs}+1},\ldots,Y_{n}$  whose covariate vectors $\bm{X}_{n_{obs}+1},\ldots,\bm{X}_n$ lie within ${\cal C}_{A}$, and we augment our recovered data with the observed data whose $p$-dimensional covariate vectors also lie within  ${\cal C}_{A}$, then only the intercept in the missing mechanism's linear predictor changes. 
In particular, the coefficients in front of terms involving $y$ in the augmented data are the same as its counterpart based on the full sample, 
permitting MNAR's presence to be reliably inferred using the augmented sample. 

Since in Corollary~\ref{cor1}, $c_2$ can take any value in the interval $(0,1]$, when considering the logit link function we will set $c_2=1$. This results in more observations in the augmented data and the variance of estimators is reduced. For any other link function, we will assume $c_2$ is obtained from Corollary~\ref{c_2_corr}.
\subsection{Importance of restricting augmented covariates to ${\cal C}_A$ }\label{sec:type_1_error}
In this section, we demonstrate the importance of restricting the augmented covariates to lie within the $p$-dimensional region ${\cal C}_A$ (the condition $\bm{x} \in {\cal C}_A$). We will do so by considering two specific forms  of $\bm{w}$ and $\bm{z}$ that will be used throughout the numerical examples of this paper. We will only consider the logit link function in this section.
\begin{itemize}
\item {\bf Scenario 1:} We set  $\bm{w} = (1,x_1,\ldots,x_p)^T$ and $\bm{z} =(y) $. In this setting, we let $\bm{\lambda}=(\alpha_0,\ldots,\alpha_{p})^T$ and $\bm{\psi}=(\alpha_{p+1})$ for arbitrary real valued $\alpha_0,\ldots,\alpha_{p+1}.$

\item {\bf Scenario 2:} In Scenario 2, we set $p=1$ in \eqref{Main_regression_model}, $\bm{\lambda}=(\alpha_0,\alpha_{1})^T$, $\bm{\psi}=(\alpha_{2},\alpha_{3})^T$ and let $\bm{w} = (1,x_1)^T$ and $\bm{z} =(x_1y,y)^T.$ I.e. we include the interaction between $x_1$ and $y$.
\end{itemize} 
For Scenario 1, application of Corollary~\ref{cor1} provides for $\bm{x} \in  {\cal C}_A$:
\be 
{\rm Pr}(M_A\!=\!1 \, \vert \, Y_A\!=\!y,\bm{X}_A \!=\! \bm{x})=\frac{\exp(\alpha_0+ \log(c^{*})+\alpha_1x_1+ \ldots +\alpha_px_p+ \alpha_{p+1}y)}{1+\exp(\alpha_0+ \log(c^{*})+\alpha_1x_1+ \ldots +\alpha_px_p+ \alpha_{p+1}y)}  \, ,
\label{scenario_one}
\ee
where $c^*$ is given in Theorem~\ref{main_theorem}.
In Scenario 2, from Corollary~\ref{cor1}  we obtain:
\be 
{\rm Pr}(M_A\!=\!1 \, \vert \, Y_A\!=\!y,\bm{X}_A \!=\! \bm{x})=\frac{\exp(\alpha_0+ \log(c^{*})+\alpha_1x_1+\alpha_{2}y+\alpha_{3}x_1y_1)}{1+\exp(\alpha_0+ \log(c^{*})+\alpha_1x_1+\alpha_{2}y+\alpha_{3}x_1y_1)}  \, .
\label{scenario_two}
\ee
If the covariates of the augmented data, $\bm{X}_A$,  are not solely restricted to ${\cal C}_A$, then the expressions in \eqref{scenario_one} and \eqref{scenario_two} do not hold. An important consequence of this 
is that under MAR we may have $\bm{\psi}_A \neq 0$. Specifically for these two scenarios,  there is the potential for dependencies to be induced between $M$ 
and $y$ in Scenario 1 and/or $M$ and $x_1y$ in Scenario 2 {\em in the augmented sample}. This may lead to an erroneous Type 1 error when performing the hypothesis test outlined in Section~\ref{sec:test} based on this subsample. This is illustrated with a numerical example below.

Let us consider both scenarios within one simulated example. Suppose we set $p=1$ and generate $n=1000$ points from the regression model
$Y|(X=x) \sim N(2-2x, 4) \,,$
with $X \sim N(0,16)$. MAR missingness is introduced into the model using: 
\bea
P(M=1|Y=y,{ X} = { x})=\exp(-2+0.4 x )/(1+\exp(-2+0.4 x)) \,.
\eea
We will explore three different constructions of the augmented data. Our first scheme will randomly recover $n^*$ missing responses (simply a random sample without replacement) and includes all the observed data when fitting the model.
This corresponds to the choice ${\cal C}_A = \mathbb{R}$ and therefore the forms of \eqref{scenario_one} and \eqref{scenario_two} hold with $c^*=c_1$ (recall for the logit model we set $c_2=1$). Our second scheme recovers the missing responses with the $n^*$ highest values of the covariate and then augments this with all of the observed data. The final scheme is similar to Scheme 2, but we recover the responses with the $n^*$ smallest values of the covariate and subsequently augment this with all of the observed data. Schemes 2 and 3, do not include the requirement for $x\in {\cal C}_A$. To obtain estimates of Type 1 errors after obtaining the augmented data, we perform the test outlined in Section~\ref{sec:test} with the appropriate model and repeat the whole procedure 10,000 times.

\begin{figure}[h]
\centering
  \includegraphics[width=.45\linewidth]{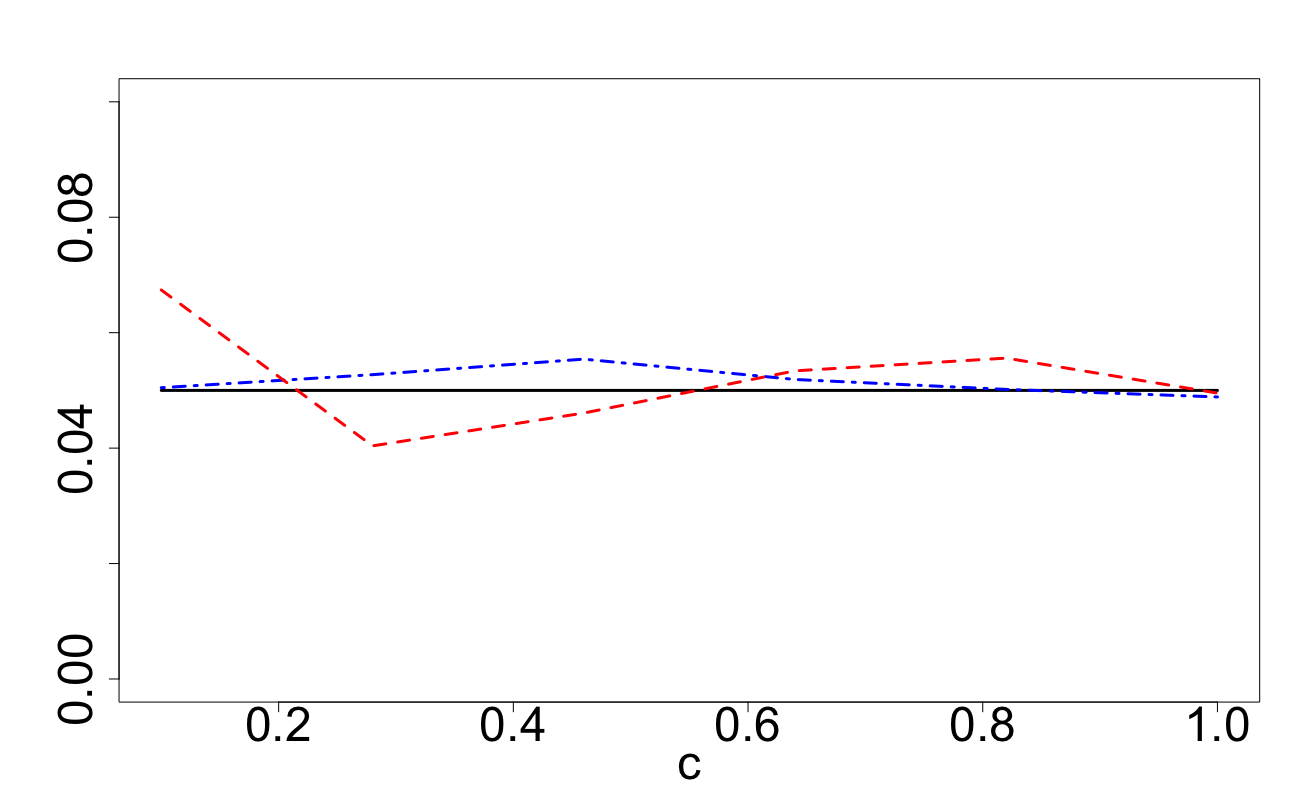}
  \includegraphics[width=.45\linewidth]{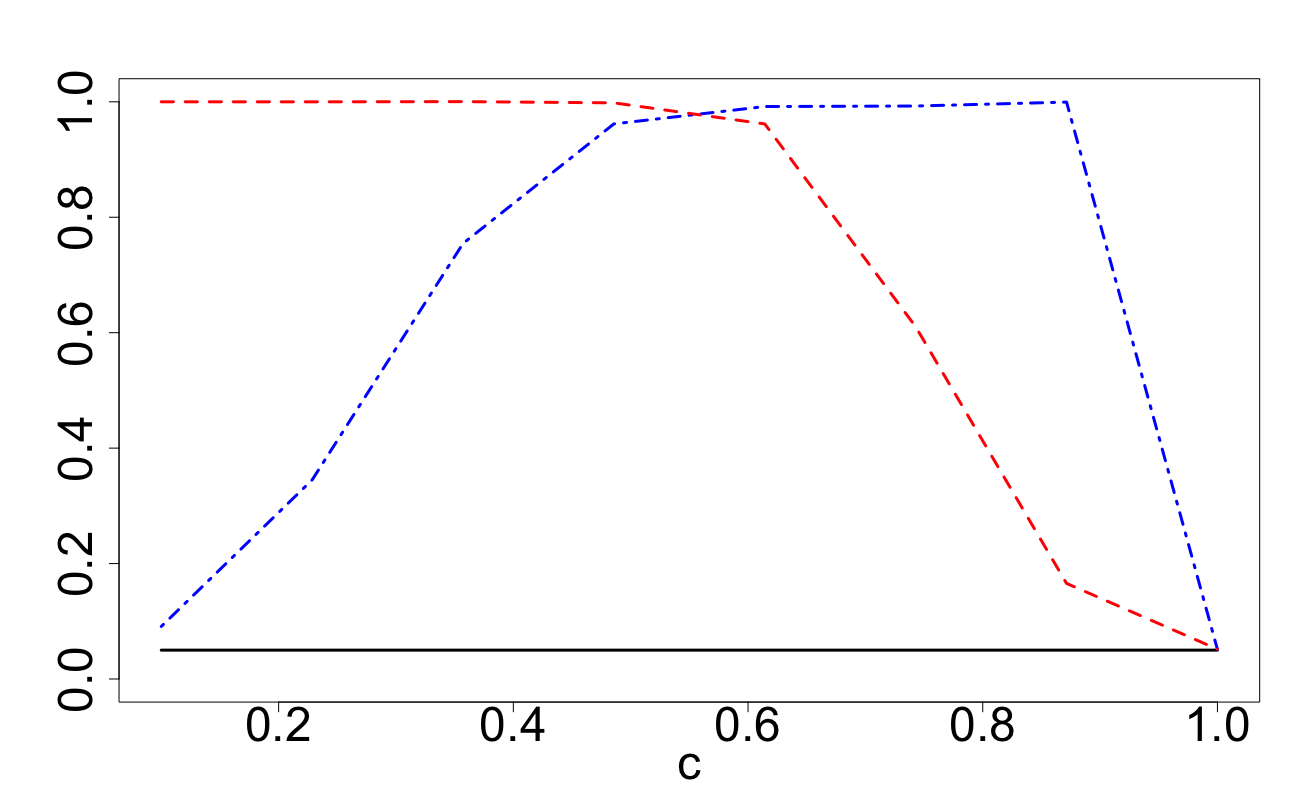}
 \caption{Empirical Type 1 error rates across different recovery proportions for Scenario 1 (L) and Scenario 2 (R) for the three recovery schemes.  
 }
 \label{type_one_error11}
\end{figure}

In Figure~\ref{type_one_error11} (L), we depict the Type 1 errors for Scenario 1, testing $H_0: \bm{\psi}_A=0$ as a function of the recovery proportion, $c=c_1$, for the three different recovery schemes. 
Similarly, Figure~\ref{type_one_error11} (R) 
presents Type I errors in Scenario 2, now testing $H_0: \bm{\psi}_A=(0,0)$. 
In Scenario 1, we see Schemes 2 and 3 exhibit modest variability around the nominal Type 1 error rate of 5\%, and the potential issue around Type 1 error inflation is not very prominent. However, the situation is vastly different in Scenario 2. 
Schemes 2 and 3 deviate significantly from 5\% here, and the use of the test cannot be recommended in these settings. This highlights the need to follow the principles derived above when constructing the recovery sample, and as expected, we see Scheme 1 results in empirical Type I error rates close to nominal
for all $c=c_1$ across both scenarios.


{The greatly inflated Type I error rates in Scenario 2 are significant, as this model (including an interaction term $x_1y$) corresponds to standard guidance on model specification to test for MNAR \citep{carpenter2012multiple}. The guidance adheres to the principle that overfitting the model is preferable to underfitting, to avoid missing relevant effects. However, naively applying this model without taking into account the properties derived above can lead to erroneously detecting the presence of MNAR with inevitable consequences for inferences. Having established conditions necessary for appropriate application of the statistical test, in the next section we derive a framework to construct recovery designs that optimizes the power of the test.}

\section{Designing the region ${{\cal C}}_{A}$}\label{sec:design}

\subsection{Approximating deviance by a non-central $\chi^2$ distribution }\label{design1}

Hereon, we operate under the conditions of Theorem~\ref{main_theorem}, Corollary~\ref{c_2_corr} and Corollary~\ref{cor1}. Assuming the logit link function, we can then take $c_2=1$ and have $\bm{\lambda}_A=\bm{\lambda}+(\log(c^*),0,\ldots,0)^T $ and $\bm{\psi}_A=\bm{\psi}$. For any other link function, we determine $c_2$ from Corollary~\ref{c_2_corr} and have $\bm{\lambda}_A=\bm{\lambda}$ and $\bm{\psi}_A=\bm{\psi}$. To proceed in maximizing the power of the likelihood ratio test for $\bm{\psi}=0$ which, as described in Section \ref{sec:test}, is equivalent to testing between MAR (Null) and MNAR (Alternative), we incorporate the findings of \cite{self1992power}. The main focus of this section is to approximate the power of the likelihood ratio test by approximating the distribution of the log-likelihood ratio statistic with a non-central chi square distribution with $s$ degrees of freedom. The noncentrality parameter used in the approximation is computed by equating the expected value of a noncentral chi-square random variable to an approximation of the expected value of the likelihood ratio statistic, 
which involves taking the expected value of lead terms in an asymptotic expansion of the likelihood ratio statistic. Specifically, the technique of \cite{self1992power} is adopted to decompose the likelihood ratio statistic into the following three components:
\be
2[ l( \bm{\hat{\psi}}, \bm{\hat{\lambda}}_A )-l( \bm{{\psi_0}}, \bm{\hat{\lambda}_0} ) ] &=& 2[ l( \bm{\hat{\psi}}, \bm{\hat{\lambda}}_A )-l( \bm{{\psi}}, \bm{{\lambda}}_A ) ]
- 2[ l( \bm{{\psi_0}}, \bm{\hat{\lambda}_0} )-l( \bm{{\psi_0}}, \bm{{\lambda^*_0}} ) ] \nonumber \\
&+&2[ l( \bm{{\psi}}, \bm{{\lambda}}_A )-l( \bm{{\psi_0}}, \bm{{\lambda^*_0}} ) ] \label{expansion_log} \,,
\ee
where $\bm{{\lambda^*_0}}$ is the limiting value of $\bm{\hat{\lambda}_0}$ and is given in Lemma~\ref{limit_null} below. The asymptotic expansion of the first component in \eqref{expansion_log} was considered in \cite{cordeiro1983improved}. Taking only the lead term in this expansion results in an approximate expected value for the first component in \eqref{expansion_log} of $q+s$. The expected value of the lead term in this expansion of the second component in \eqref{expansion_log} is equal to the trace of the matrix $A$ given by
\bea
A= \left \{  \mathbb{E}\left[- \frac{\partial^2 l(\bm{\psi},\bm{\lambda}_A)}{\partial \bm{\lambda}_A^2} \bigg\vert_{(\bm{\psi_0},\bm{\lambda^*_0})}  \right]  \right\}^{-1} \mathbb{E} \left [ \frac{\partial l (\bm{\psi},\bm{\lambda}_A)}{\partial \bm{\lambda}_A}\bigg\vert_{(\bm{\psi_0},\bm{\lambda^*_0})}  \right ]^{\otimes2} \,.
\eea
Here for a vector $\bm{a}$, $\bm{a}^{\otimes2} = \bm{a}\bm{a}^T$. In \cite{self1992power}, formulae for computing $A$ are provided for generalized linear models. Here, we will derive the results for the specific case of a logit model, but other link functions can easily be considered. We obtain from \citet[p. 33]{self1992power}:
\bea
\mathbb{E}\left[- \frac{\partial^2 l(\bm{\psi},\bm{\lambda}_A)}{\partial \bm{\lambda}_A^2} \bigg\vert_{(\bm{\psi_0},\bm{\lambda^*_0})}  \right]  = \sum_{i=1}^{n_A} \frac{\exp(\theta^*_i)}{(1+\exp(\theta^*_i))^2}\bm{w}_i^{\otimes2} \,\,\, 
\eea
\text{ and }
\bea \,\, \mathbb{E} \left [ \frac{\partial l (\bm{\psi},\bm{\lambda}_A)}{\partial \bm{\lambda}_A} \bigg\vert_{(\bm{\psi_0},\bm{\lambda^*_0})} \right ]^{\otimes2} =  \sum_{i=1}^{n_A} \frac{\exp(\theta_i)}{(1+\exp(\theta_i))^2}\bm{w}_i^{\otimes2} \,,
\eea
where $\theta^*_i = \bm{w}_i^T\bm{{\lambda^*_0}}$ and $\theta_i  =  \bm{w}_i^T\bm{\lambda}_A+ \bm{z}_i^T\bm{\psi}$.\\
The expected value of the third component in \eqref{expansion_log}, denoted by $\Delta$, can be calculated exactly. For the logit link function, it is given by 
\bea
\Delta = 2\sum_{i=1}^{n_A} \left \{ \frac{\exp(\theta_i)}{1+\exp(\theta_i)}(\theta_i-\theta^*_i) - \log \left(\frac{1+\exp(\theta_i)}{1+\exp(\theta^*_i)} \right) \right \} \,.
\eea
Since the expected value of a noncentral chi-square random variable with $s$ degrees of freedom and noncentrality parameter $\gamma$ is $s+\gamma$, using the recommendation of \cite{self1992power}, we approximate the distribution of the likelihood ratio statistic by a noncentral chi-square distribution with noncentrality parameter $q+\Delta-tr(A)$.

To compute $A$ and $\Delta$, one has to know the limiting value of $\bm{\hat{\lambda}_0}$.  For an arbitrary link function $g$, we can appeal to the following lemma.
\begin{lemma}\label{limit_null}
The limiting value of $\bm{\hat{\lambda}_0}$, that is $\bm{{\lambda^*_0}}$, minimizes the Kullback-Leibler divergence between the alternative and null models assuming the alternative model is true. Consequently, it satisfies:
\bea
\arg \min_{\bm{{\lambda^*_0}}} \mathbb{E}_1 \bigg [&&  \!\!\!\!\!\!\!\!\!\!\!\!\!\! g^{-1}(\bm{W}^T\bm{\lambda_A}+\bm{Z}^T\bm{\psi})\log\left ( \frac{ g^{-1}(\bm{W}^T\bm{\lambda_A}+\bm{Z}^T\bm{\psi})}{ g^{-1}(\bm{W}^T\bm{{\lambda^*_0}})} \right )  \\ 
&+& (1- g^{-1}(\bm{W}^T\bm{\lambda_A}+\bm{Z}^T\bm{\psi}))\log\left( \frac{1- g^{-1}(\bm{W}^T\bm{\lambda_A}+\bm{Z}^T\bm{\psi})}{1- g^{-1}(\bm{W}^T\bm{{\lambda^*_0}})} \right)  \bigg] \,,
\eea

where the capitalization $\bm{W}$ and $\bm{Z}$ emphasizes that the elements of $\bm{w}$ and $\bm{z}$ are now functions of the random variables defined in Lemma~\ref{mixture_formulation}, $\mathbb{E}_1$ denotes expectation under the alternative model and is taken with respect to $\bm{X}_A$ and ${Y}_A$.
\end{lemma}

To compute the expectation with respect to $\bm{X}_A$ and $Y_A$, one can exploit the forms of \eqref{key_mixture} and \eqref{key_mixture2}. After conditioning on $M_A=1$ and $M_A=0$, one needs to know the joint distributions of $(\bm{X}_R,Y_R)$ and $(\bm{X}_O,Y_O)$ respectively to compute the required expectation. 

\begin{lemma} \label{auxillary_lem1}
For $\bm{x} \in {\bf {\cal C}}_{A}$:
\begin{eqnarray*}
{\rm Pr}(\bm{X}_R\in \bm{dx},Y_R \in dy) &=& \frac{{\rm Pr}(M=1 \,\vert\, \bm{X}=\bm{x},Y = y){\rm Pr}(Y\in dy \, \vert \, \bm{X}=\bm{x}){\rm Pr}(\bm{X}\in \bm{dx})}{{\rm Pr}({\cal M}_R )} \\
{\rm Pr}(\bm{X}_O\in \bm{dx},Y_O \in dy) &=& \frac{{\rm Pr}(M=0 \,\vert \,\bm{X}=\bm{x},Y = y){\rm Pr}(Y\in dy \,\vert \, \bm{X}=\bm{x}){\rm Pr}(\bm{X}\in \bm{dx})}{{\rm Pr}({\cal M}_O )} \,,
\end{eqnarray*}
otherwise zero. Here, $\bm{dx}=(dx_1,\ldots,dx_p)$ is a vector of infinitesimals and the relation ${{\rm Pr}(Y\in dy \, \vert \, \bm{X}=\bm{x})}$ can be obtained from \eqref{Main_regression_model}. 
\end{lemma}
From Lemma~\ref{auxillary_lem1}, it is clear we need to have knowledge of the initial missing mechanism ${\rm Pr}(M=1 \,\vert \,\bm{X}=\bm{x},Y = y)$, the regression relation  ${\rm Pr}(Y\in dy \,\vert \, \bm{X}=\bm{x})$ and the distribution of $\bm{X}$ to compute expectations. Whilst this appears restrictive, we find, through an empirical evaluation, our results are robust to various types of misspecification; see Section~\ref{sec:robustness}.

The matrix $A$ and scalar $\Delta$ are random quantities since they depend on $\bm{w}_i$. However, by the law of large numbers, for the logit model as ${n_A\rightarrow \infty}:$
\bea
&&\frac{1}{n_A}\sum_{i=1}^{n_A} \frac{\exp(\theta^*_i)}{(1+\exp(\theta^*_i))^2}\bm{w}_i^{\otimes2} \rightarrow \mathbb{E} \left[ \frac{\exp(\theta^*)}{(1+\exp(\theta^*))^2}\bm{W}^{\otimes2}  \right]\\
&&\frac{1}{n_A}\sum_{i=1}^{n_A} \frac{\exp(\theta_i)}{(1+\exp(\theta_i))^2}\bm{w}_i^{\otimes2} \rightarrow \mathbb{E} \left[ \frac{\exp(\theta)}{(1+\exp(\theta))^2}\bm{W}^{\otimes2}  \right]\,,
\eea
where $\theta^* =\bm{W}^T\bm{{\lambda^*_0}}$ and $\theta=\bm{W}^T\bm{\lambda}_A+ \bm{Z}^T\bm{\psi}$.

Therefore, for $n_A$ suitably large, we propose replacing each term in $A$ with their limiting forms. We also suggest replacing $\Delta$ with $\mathbb{E}\Delta$ (although this expectation will be dropped from notation). The expectations can be computed by averaging over $\bm{X}_A$ using Lemma~\ref{auxillary_lem1}.

We now propose two similar algorithms for choosing our recovery design. The first algorithm will be aimed at GLMs with an arbitrary link function whereas the second will focus only on the case of the logit link, which has the appealing property alluded to in Corollary~\ref{cor1}.
For simplicity, we only consider ${\bf {\cal C}}_{A}$ taking the form of a $p$-dimensional cuboid.
\begin{itemize}
\item {\bf Algorithm 1a (general link function).}
For a given $0<c_1\leq 1$, determine $c_2$ from Corollary~\ref{c_2_corr}. Then select the $p$-dimensional cuboid ${\bf {\cal C}}_{A}$ such that the noncentrality parameter $q+\Delta-tr(A)$ is maximized subject to the constraint ${\rm Pr}({\cal M}_R)\geq c_1\cdot {\rm Pr}(M=1)$ and ${\rm Pr}({\cal M}_O)>0$. The recovery design ${\bf {R}}$ should consist of a random sample of $n^*$ points within ${\bf {\cal C}}_{A}$ and should be augmented {\em with a random sample containing $c_2\times100\%$ of the observed data} within ${\bf {\cal C}}_{A}$.

\item {\bf Algorithm 1b (logit link function).}
For a given $0<c_1\leq 1$, set $c_2=1$ and select the $p$-dimensional cuboid ${\bf {\cal C}}_{A}$ such that the noncentrality parameter $q+\Delta-tr(A)$ is maximized subject to the constraint ${\rm Pr}({\cal M}_R)\geq c_1\cdot {\rm Pr}(M=1)$ and ${\rm Pr}({\cal M}_O)>0$. The recovery design ${\bf {R}}$ should consist of a random sample of $n^*$ points within ${\bf {\cal C}}_{A}$ and should be augmented {\em with all of the observed data} within ${\bf {\cal C}}_{A}$.
\end{itemize}

In practice, implementation of Algorithm 1 (in both cases) is likely to lead to locally optimal solutions. Furthermore, in a finite regime there will likely be scenarios where ${\bf {\cal C}}_{A}$ includes slightly fewer points than $n^*$. In this case, we recommend uniform enlargement until $ n^* $ points lie within ${\bf {\cal C}}_{A}$. In Section~\ref{sec:simulation}, we investigate the performance of recovery designs found by Algorithm~1, measured by the corresponding power of the test, and compare them to the benchmark random design (described by Scheme 1 in Section \ref{sec:type_1_error}).

\subsection{Minimizing asymptotic variance}\label{design2}

In this section, we briefly explore an alternative approach to design the region ${ {\cal C}}_{A}$, and thus the recovery design, to try to optimize the power of the SMF test \eqref{SMF_test}. This method is more in line with classical design theory, in particular $D_1$-optimality. {However, the 
derivations here only permit scalar $\bm{\psi}$. The most relevant setting that satisfies this condition is described by Scenario 1 in Section~\ref{sec:type_1_error}, where only the main effect of $y$ is included in the SMF test. Accordingly, we only consider $\bm{w}$ and $\bm{z}$ of the form $\bm{w} = (1,x_1,\ldots,x_p)^T$ and $\bm{z} =(y) $.} Note, under regularity conditions, the MLE of $\bm{\alpha}:=(\bm{\lambda}_A,\bm{\psi})^T$ satisfies 
\begin{eqnarray}
( \bm{\hat{\alpha}}- \bm{\alpha})\stackrel{d}{\rightarrow} N({\bm 0},I^{-1}({\bm\alpha})) \,\,\,\text{ as }\,\, n_A\rightarrow \infty\,,  \label{classic_MLE}
\end{eqnarray}
where $n_A$ is the number of observations in the augmented data 
and $I(\bm{\alpha})$ is the observed Fisher Information matrix of $\bm{\alpha}$. A design that provides the most precise information about $\alpha_{p+1}$, the coefficient of $y$, 
is one that minimizes the (asymptotic) variance of ${\hat{\alpha}_{p+1}}$ or, equivalently, $I^{-1}({\bm\alpha})[p+1,p+1]$, the $(p+1)$th diagonal element of $I^{-1}({\bm\alpha})$. \cite{atkinson1975design} showed that, for completely observed data, this approach is equivalent to maximizing the asymptotic power of the test that the coefficient of interest is zero. 

We avoid inverting $I(\bm{\alpha})$ 
by appealing to Cramer's rule to obtain
\begin{eqnarray}\label{asymptotic_var}
 I^{-1}({\bm\alpha})[p+1,p+1]= \mbox{det}(I(\bm{\alpha})[1:p,1:p])/\mbox{det}(I(\bm{\alpha}))  \,,
\end{eqnarray}
where $I(\bm{\alpha})[1:p,1:p]$ is the $p\times p$ submatrix of $I(\bm{\alpha})$ comprising its first $p$ rows and columns.
$I(\bm{\alpha})$ satisfies $I(\bm{\alpha}) = -\mathbb{E}[H(l(\bm{\alpha}))]$ where $H$ is the Hessian and $l$ is the log-likelihood function of the augmented data. By considering missing mechanisms of the form \eqref{expit_model}, its properties (continuity, differentiability, and being bounded on $(0,1)$) mean we can take the expectation within the Hessian. Exploiting i.i.d. observations,
we obtain
\begin{eqnarray}
I(\bm{\alpha}) 
&=&\!-H \big(  \mathbb{E}[n_A]\!\cdot\!\mathbb{E} \big( M_A \log \!\left[
g^{-1}(\bm{W}^T\bm{\lambda}_A + \bm{Z}^T\bm{\psi}_A) \right ]  \nonumber 
\\
&& + (1-M_A) \log \!\left[1-g^{-1}(\bm{W}^T\bm{\lambda}_A + \bm{Z}^T\bm{\psi}_A) \right]  \big)\! \big) \,.  \nonumber \label{asymptotic_likelihood}
\end{eqnarray}
One can also show that $ \mathbb{E}[n_A] = n \cdot (c_1{\rm Pr}(M=1)+c_2{\rm Pr}({\cal M}_O ))$.

For the logit link function, we have $\bm{\lambda}_A=\bm{\lambda}+(\log(c^*),0,\ldots,0)^T $ and $\bm{\psi}_A=\bm{\psi}$. Otherwise, provided we operate under the conditions of Corollary~\ref{c_2_corr} and choose $c_2$ accordingly, we have for any other link function, $\bm{\lambda}_A=\bm{\lambda}$ and $\bm{\psi}_A=\bm{\psi}$. To compute expectations in \eqref{asymptotic_likelihood}, we appeal to Lemma~\ref{auxillary_lem1}.

We now propose an alternative algorithm to Algorithm 1. 
Again, for simplicity, we only consider ${\bf {\cal C}}_{A}$ taking the form of a $p$-dimensional cuboid. 
\begin{itemize}
\item {\bf Algorithm 2a (arbitrary link function):}
For a given $0<c_1\leq 1$, determine $c_2$ from Corollary~\ref{c_2_corr}. Then select the $p$-dimensional cuboid ${\bf {\cal C}}_{A}$ such that the r.h.s of \eqref{asymptotic_var} is minimized subject to the constraint  ${\rm Pr}({\cal M}_R)\geq c_1\cdot {\rm Pr}(M=1)$ and ${\rm Pr}({\cal M}_O)>0$. The recovery design ${\bf {R}}$ comprises a random sample of $n^*$ points within ${\bf {\cal C}}_{A}$ augmented {\em with a random sample of $c_2\times100\%$ of the observed data} within ${\bf {\cal C}}_{A}$.

\item {\bf Algorithm 2b (logit link function):}
For a given $0<c_1\leq 1$, set $c_2=1$ and select the $p$-dimensional cuboid ${\bf {\cal C}}_{A}$ such that the r.h.s of \eqref{asymptotic_var} is minimized subject to ${\rm Pr}({\cal M}_R)\geq c_1\cdot {\rm Pr}(M=1)$ and ${\rm Pr}({\cal M}_O)>0$. The design ${\bf {R}}$ comprises a random sample of $n^*$ points within ${\bf {\cal C}}_{A}$ augmented {\em with all the observed data} within ${\bf {\cal C}}_{A}$.
\end{itemize}


\section{Simulation studies}\label{sec:simulation}

In this section, we perform a simulation study assessing the effects Algorithm 1 and Algorithm 2 have on the power of the SMF test for MNAR outlined in Section~\ref{sec:test}. In Section \ref{sec:sim1}, we solely consider Scenario 1 described in Section \ref{sec:type_1_error}), i.e. we set $\bm{w} = (1,x_1,\ldots,x_p)^T$ and $\bm{z} =(y)$, considering examples for both $p=1$ and $p=2$. In Section \ref{sec:sim2}, we consider Scenario 2 for $p=1$, i.e. additionally include an interaction term between $x_1$ and $y$ so that $\bm{w} = (1,x_1)^T$ and $\bm{z} =(x_1y,y)^T.$ 

\subsection{Scenario one: simulation study for MSE and power}\label{sec:sim1}

\begin{figure}[h]
\centering
  \includegraphics[width=0.45\linewidth]{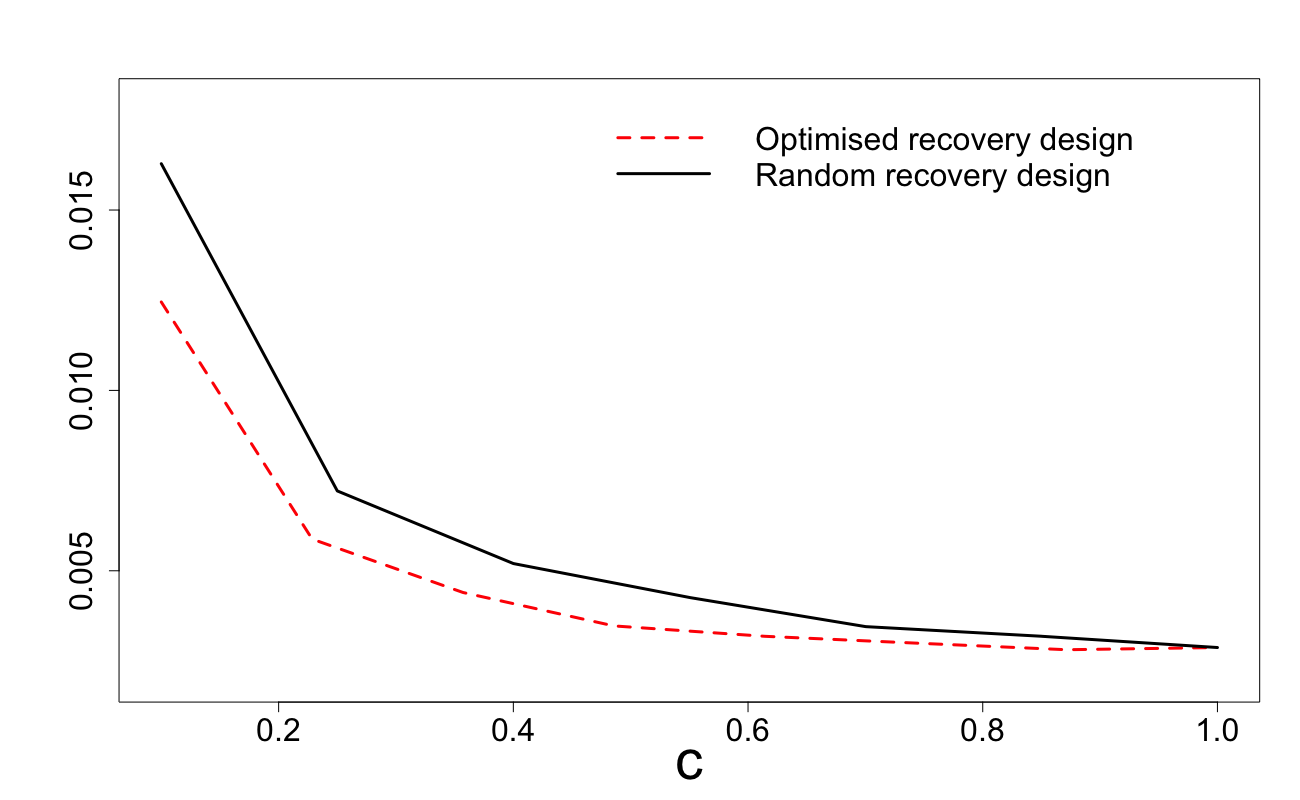}
  \includegraphics[width=0.45\linewidth]{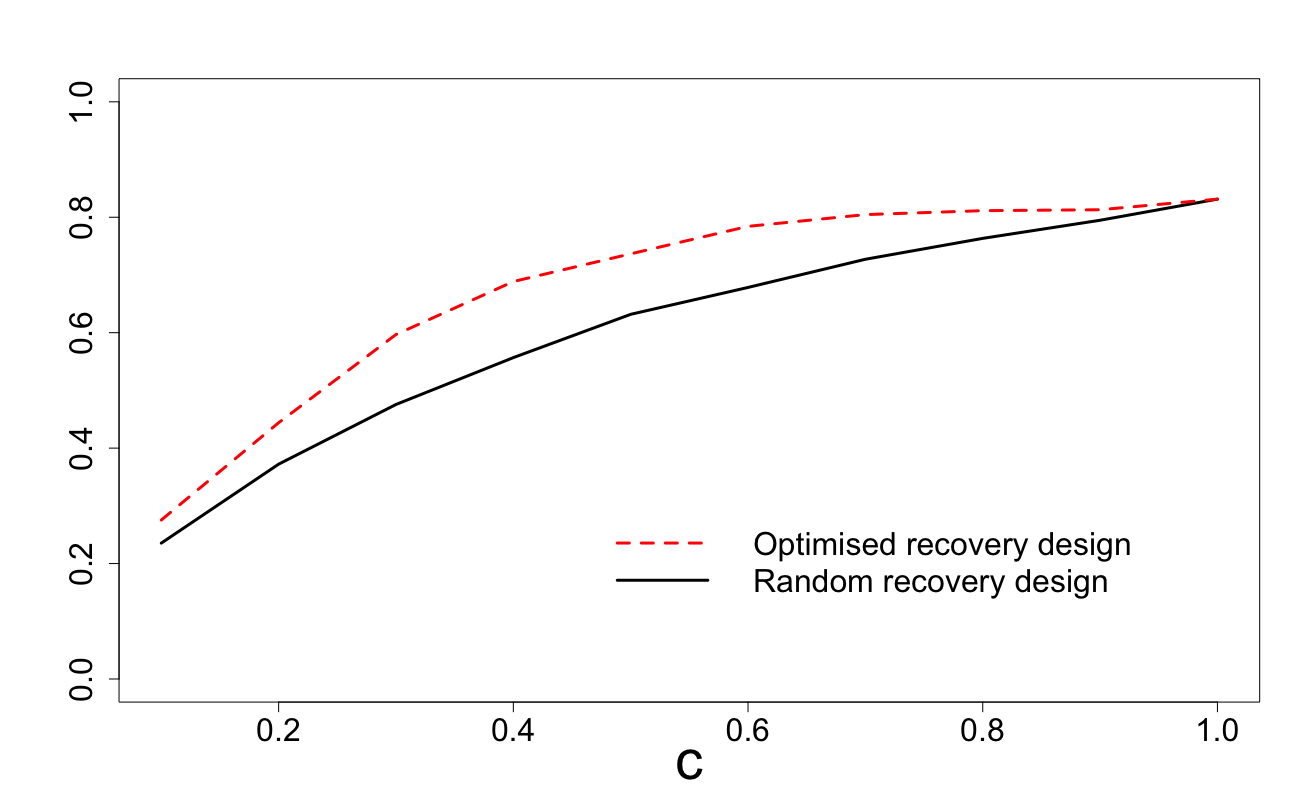}
  
  \caption{MSE (L) and Power (R) for different recovery proportions and designs; 
  $p=1$.}
    \label{fig2}
\end{figure}

{\bf Example 1.} We first consider $p=1$ (single covariate). We generate $n=1000$ points by, 
\be\label{regression_model_1}
Y|(X=x) \sim N(2-2x, 4) \,,
\ee
with $X \sim N(0,16)$.  MNAR missingness is introduced into $y$ values using 
\be\label{first_logistic_example}
{\rm Pr}(M=1|Y=y,{ X} = { x})=\exp(-2+0.4 x -0.15 y)/(1+\exp(-2+0.4 x-0.15 y)) \,
\ee
so that approximately 30\%  of points are missing their $y$ value. We repeat this process $10,000$ times and in each replication apply test \eqref{SMF_test}, 
with $H_0: \bm{\psi} = 0$, to the generated sample. 
Figure~\ref{fig2} presents the mean squared error (MSE) of the MLE, $\hat{\bm{\psi}}$, as well as the power of test \eqref{SMF_test}, for different recovery proportions, $c_1$, under designs constructed from Algorithm~2b assuming a logit link. 
The results of Algorithm~1b are indistinguishable from those of Algorithm 2b and so are not presented here. We also note that Algorithm~2 is slightly simpler to execute than Algorithm 1 for this example. We compare these to equivalent results obtained from
randomly recovered observations, 
i.e. ${\bf {\cal C}}_{A}=\mathbb{R}$. 
We see considerable gains over random sampling, in both MSE and power, when using designs found by Algorithm 2. E.g., 
when $c_1=0.3$, the power of test \eqref{SMF_test} is $\approx$ 60\% for our method while only $\approx$ 45\% for random sampling. Moreover, in order to achieve a power of 60\% with random sampling, we would need to recover almost half of the missing responses instead of only 30\% with the optimal design.

\begin{figure}[h]
\centering
  \includegraphics[width=0.45\linewidth]{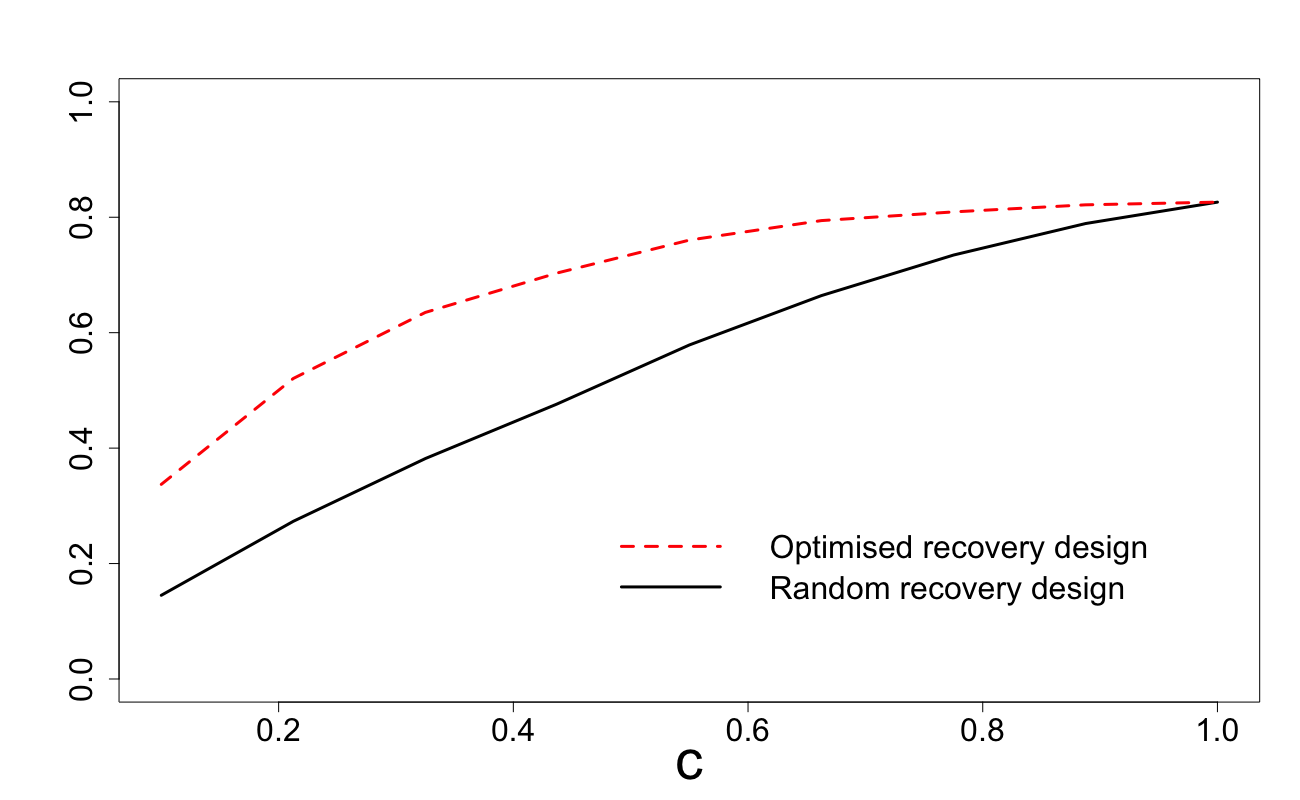}
  \includegraphics[width=0.45\linewidth]{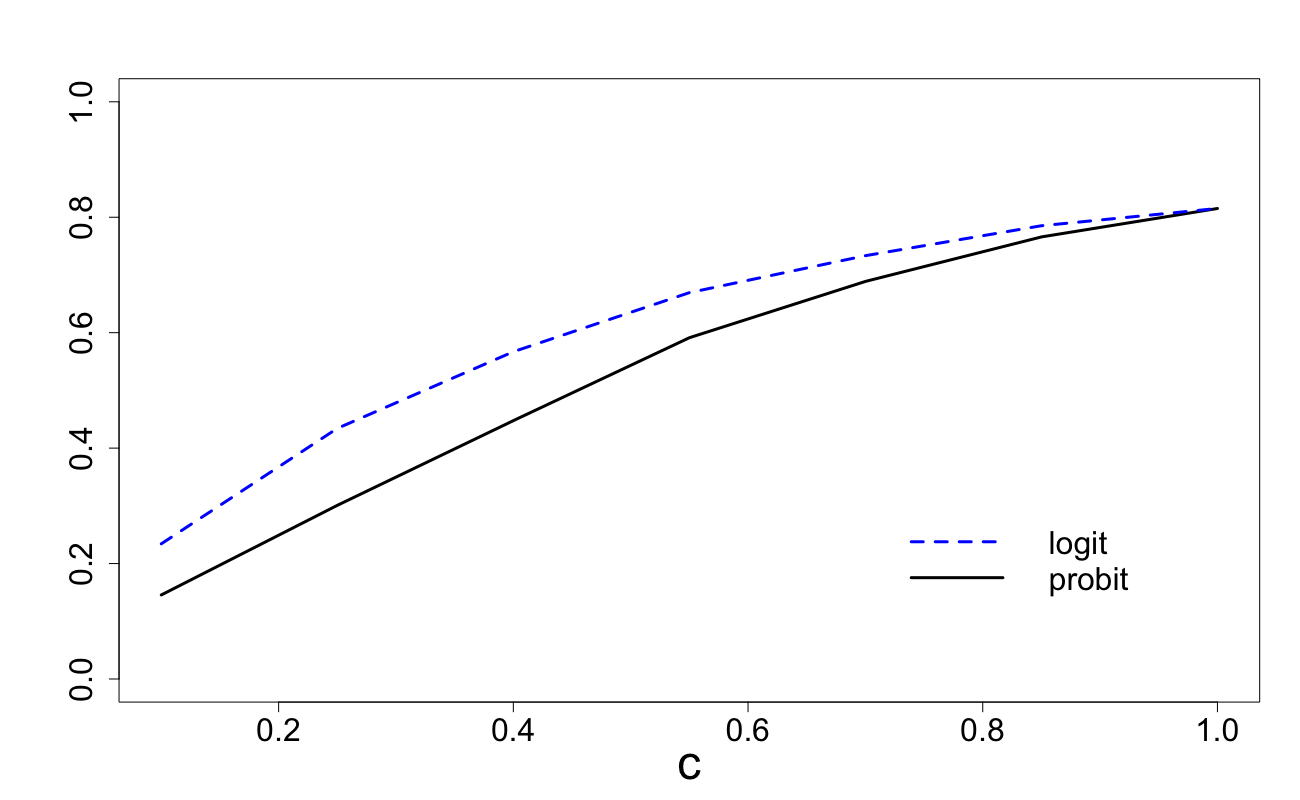}
  
  \caption{Power (L) for different recovery designs with MNAR mechanism \eqref{probit_relation}; Power (R) only for random recovery designs with MNAR mechanisms \eqref{first_logistic_example} and \eqref{probit_relation}. 
 }
    \label{probit_figures}
\end{figure}

To study the effectiveness of Algorithm 2a,  MNAR missingness is instead introduced into $y$ 
using the probit relation: 
\be \label{probit_relation}
{\rm Pr}(M=1|Y=y,{ X} = { x})= \Phi(-1.14+0.23 x -0.09 y) \,.
\ee
The choice of these parameters is to ensure the probability to be missing is approximately equal to the probability given in \eqref{first_logistic_example}, {resulting again in approximately 30\% of points with missing $y$ values}. In Figure~\ref{probit_figures} (left), we present the power of test \eqref{SMF_test} for different $c_1$ under recovery designs constructed from Algorithm~2a as well as using random recovery. We see the optimized recovery design produces significant improvements for this probit example. In Figure~\ref{probit_figures} (right), we compare the power of the random recovery designs assuming a logit link versus a probit link under \eqref{first_logistic_example} and \eqref{probit_relation} respectively. Since the parameters in both mechanisms are chosen such that the conditional probabilities $P(M=1|Y=y,{ X} = { x})$ are similar, the higher power for the logit model is due to being able to take $c_2=1$. For 
the data generated using \eqref{probit_relation}, if we fit the probit model and take $c_2=1$, instead of determining the proportion using
Theorem~\ref{main_theorem}, the resulting power is close to that obtained from the logit model in Figure~\ref{probit_figures} (right).
However, the estimates returned have little relation to the true parameters in the mechanism (except when $c_1=1$) as seen in Table~\ref{probit_table}, where for different values of $c_1$ we return estimates for $\bm{\lambda}=(\alpha_0,\alpha_1)$ and $\bm{\psi}=(\alpha_2)$ fixing $c_2=1$ and assuming ${\rm Pr}(M_A\!=\!1 \, \vert \, Y_A\!=\!y,\bm{X}_A \!=\! \bm{x})=\Phi(\alpha_0+\alpha_1x+\alpha_2y)$. 

	\begin{table}[h!]
			\caption{ Parameters estimates in \eqref{probit_relation} with $c_2=1$: True $(\alpha_0, \alpha_1, \alpha_2) = (-1.14, 0.23, -0.09$)}
		\centering
		\resizebox{15cm}{!}{\begin{tabular}{crrrrrrrrrrr}
			\hline
 Parameter & 0.1 & 0.2 & 0.3 & 0.4 & 0.5 & 0.6 & 0.7 & 0.8 & 0.9&1.0 \\ 
	  \hline
 $\alpha_0$   &  -2.244   & -1.938   & -1.751    & -1.614    &   -1.504   &  -1.407     & -1.329   & -1.261   & -1.199  & -1.143  \\
    $\alpha_1 $     & 0.199   &     0.207  &    0.214   &  0.219    &    0.222  &   0.222    &  0.225  &    0.227   &  0.228 &   0.229  \\
     $\alpha_2$         &  -0.072  &  -0.078   &    -0.080   &  -0.081   &   -0.083 &   -0.084    &  -0.084  &  -0.084  &   -0.086  &  -0.085  \\
  \hline
		\end{tabular}}
		\label{probit_table}

	\end{table}

{\bf Example 2.} We also consider $p=2$, specifically we generate $n=1000$ points by
        $ Y|(X_1=x_1,X_2=x_2) \sim N(2-2x_1+2x_2, 4)$, with $X_1 \sim N(0,16)$ and $X_2 \sim N(2,4)$. MNAR missingness is introduced into $y$ using: 
\be\label{MAR_model}
P(M=1|Y=y,{ X_1} = { x_1},X_2=x_2)=\frac{\exp(-2+0.4 x_1+0.2 x_2-0.15 y)}{1+\exp(-2+0.4 x_1+0.2 x_2-0.15 y)} \,.
\ee
{so that again approximately 30\% of points are missing $y$.} 

\begin{figure}[h]
\centering
  \includegraphics[width=0.45\linewidth]{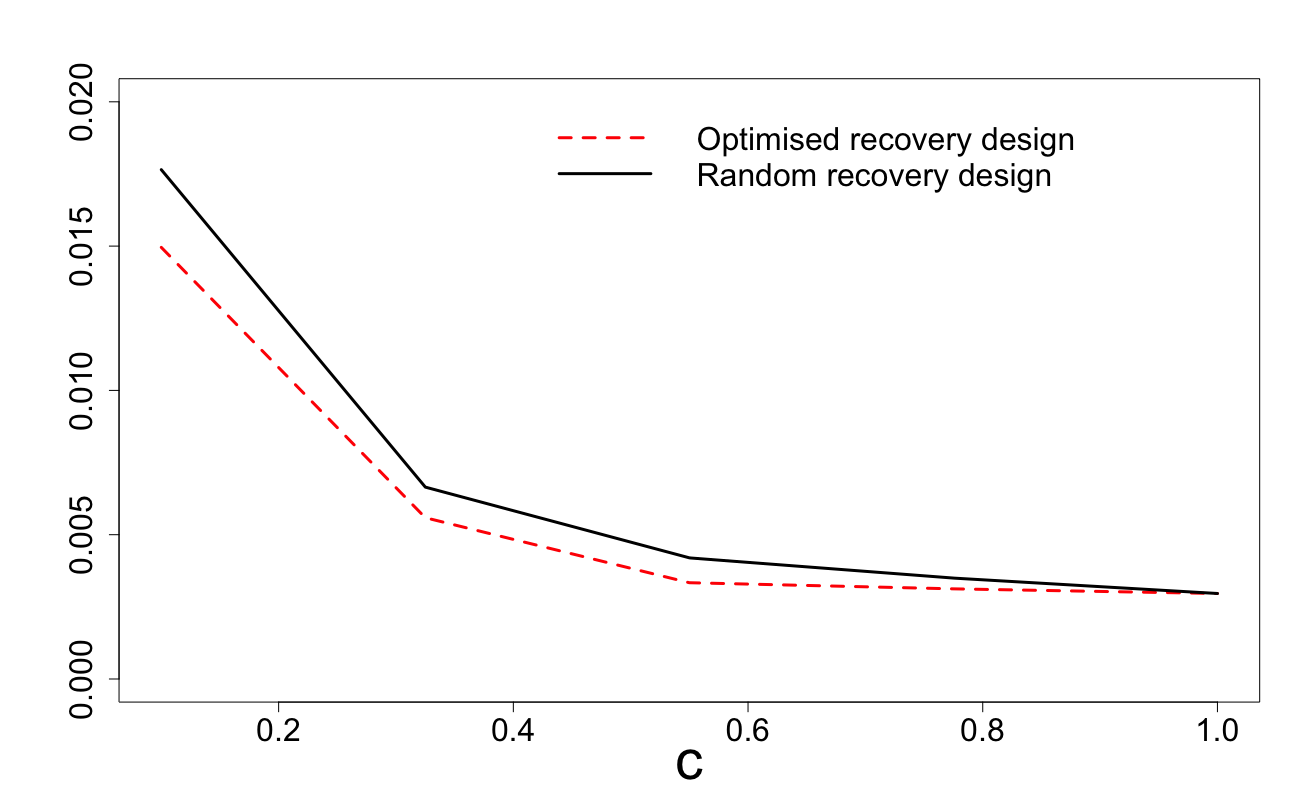}
  \includegraphics[width=0.45\linewidth]{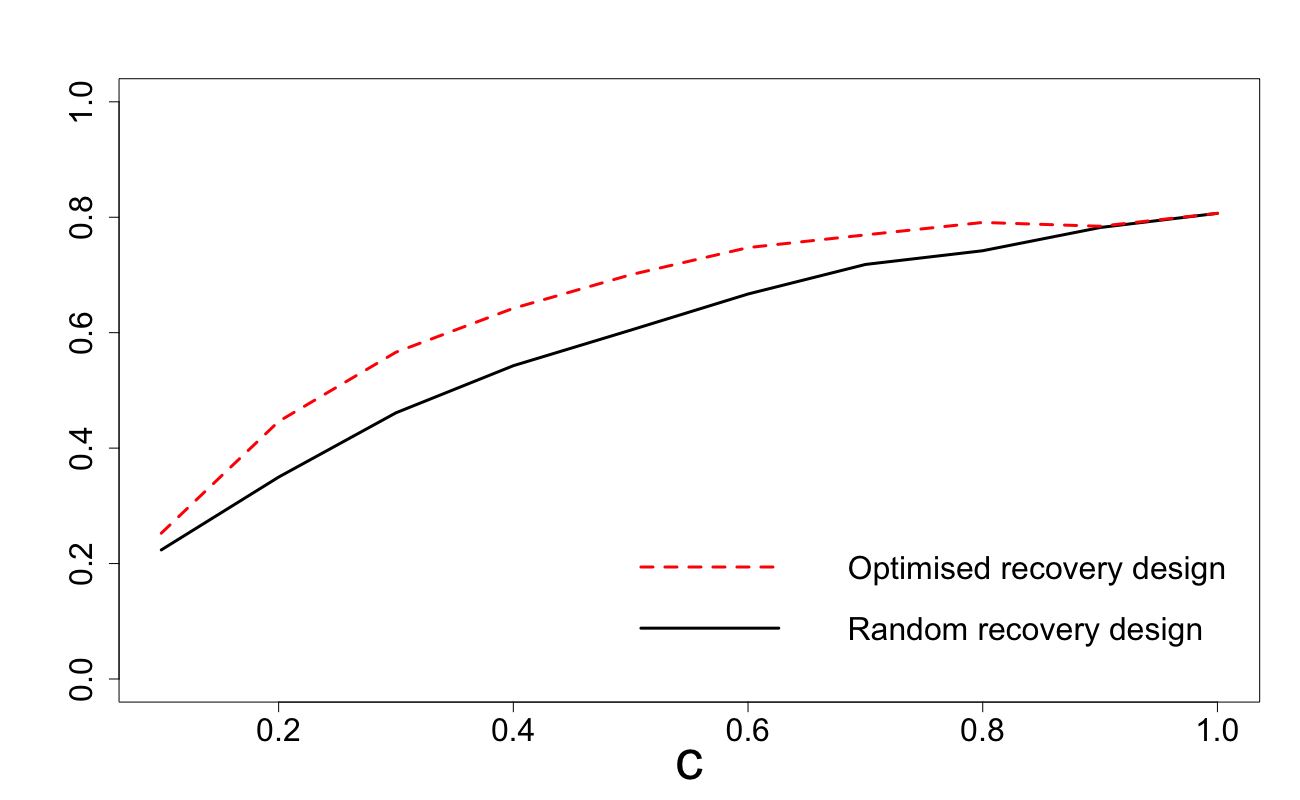}
  \caption{MSE (L) and Power (R) for different recovery designs, now with $p=2$.}
    \label{fig4}
\end{figure}

\begin{figure}[!h]
\centering
  \includegraphics[width=0.45\linewidth]{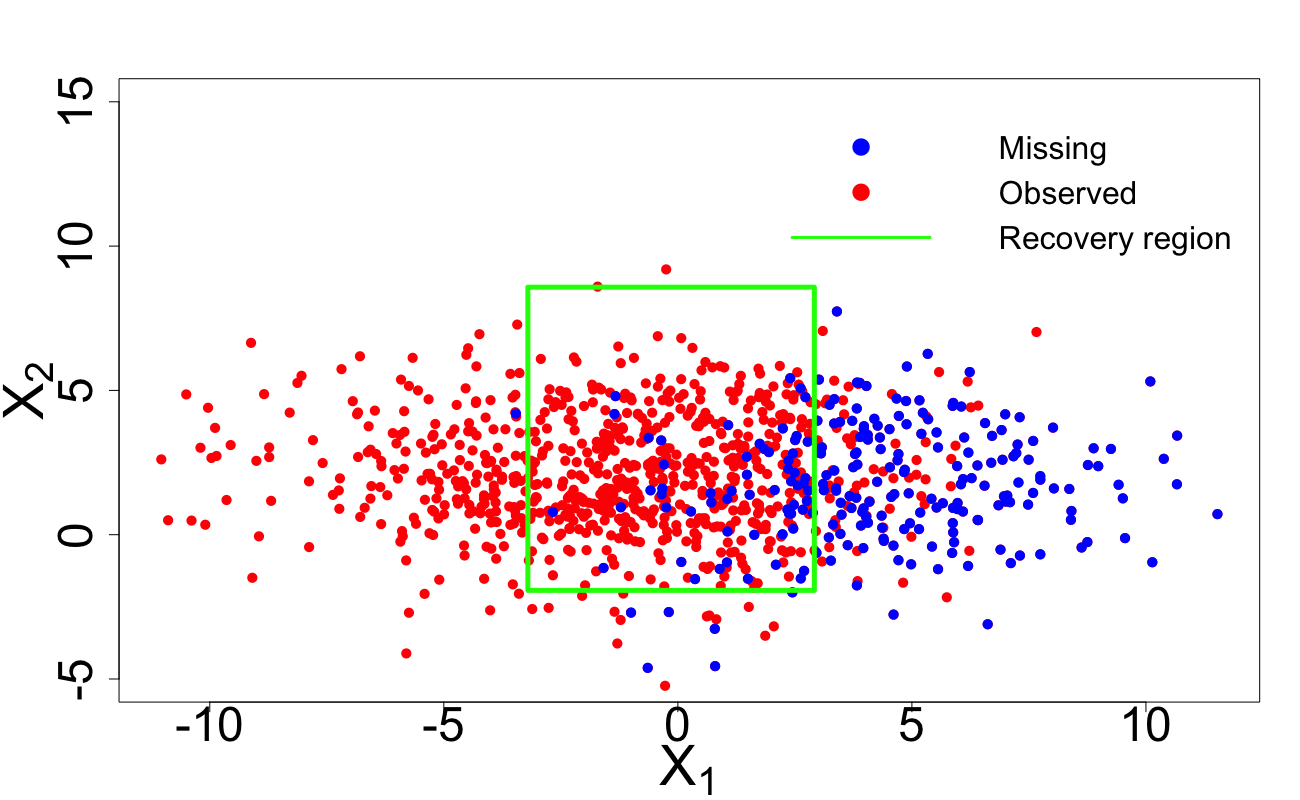}
  \includegraphics[width=0.45\linewidth]{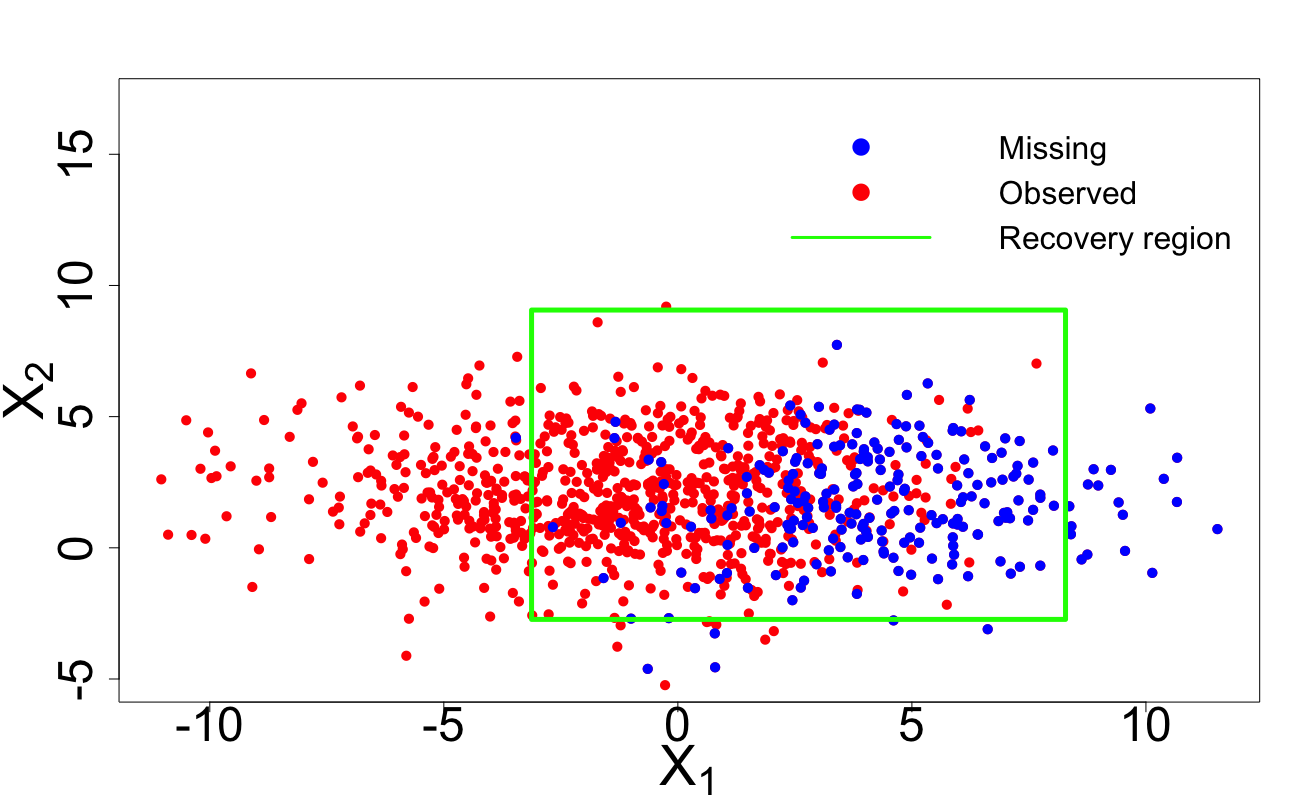}
    \caption{Recovery regions ${\bf {\cal C}}_{A}$ for $p=2$ and $c_1=0.2$ (L) and $p=2$ and $c_1=0.9$ (R). }
    \label{fig6}
\end{figure}

Figure \ref{fig4} presents equivalent summaries to Figure \ref{fig2} with similar conclusions as for Example 1. Again, Algorithm 1b and 2b produce almost identical results. Figure~\ref{fig6} illustrates the regions ${\bf {\cal C}}_{A}$, denoted by rectangles, found by Algorithm~1b for $c_1=0.2$ and $0.9$.  

\subsection{Scenario two: simulation study for MSE and power}\label{sec:sim2}
In this section, we assess the performance of Algorithm 1b since Algorithm 2b is not applicable.
We generate $n=1000$ points as follows: $X \sim N(1,4)$ with ${Y|(X=x) \sim N(2-0.5x, 4)}$. We study the performance of the algorithm for two MNAR mechanisms:\\
{\bf Example 3.}  MNAR missingness is introduced using: 
\bea
P(M=1|Y=y,{ X_1} = { x_1})=\frac{\exp(-1-0.5 x_1+0.1y +0.05 x_1y)}{1+\exp(-1-0.5 x_1+0.1y +0.05 x_1y)} \,.
\eea
{\bf Example 4.} MNAR missingness is introduced using: 
\bea
P(M=1|Y=y,{ X_1} = { x_1})=\frac{\exp(-2+0.5 x_1+0.04y +0.03 x_1y)}{1+\exp(-2+0.5 x_1+0.04y +0.03 x_1y)} \,.
\eea

\begin{figure}[H]
\centering
  \includegraphics[width=0.45\linewidth]{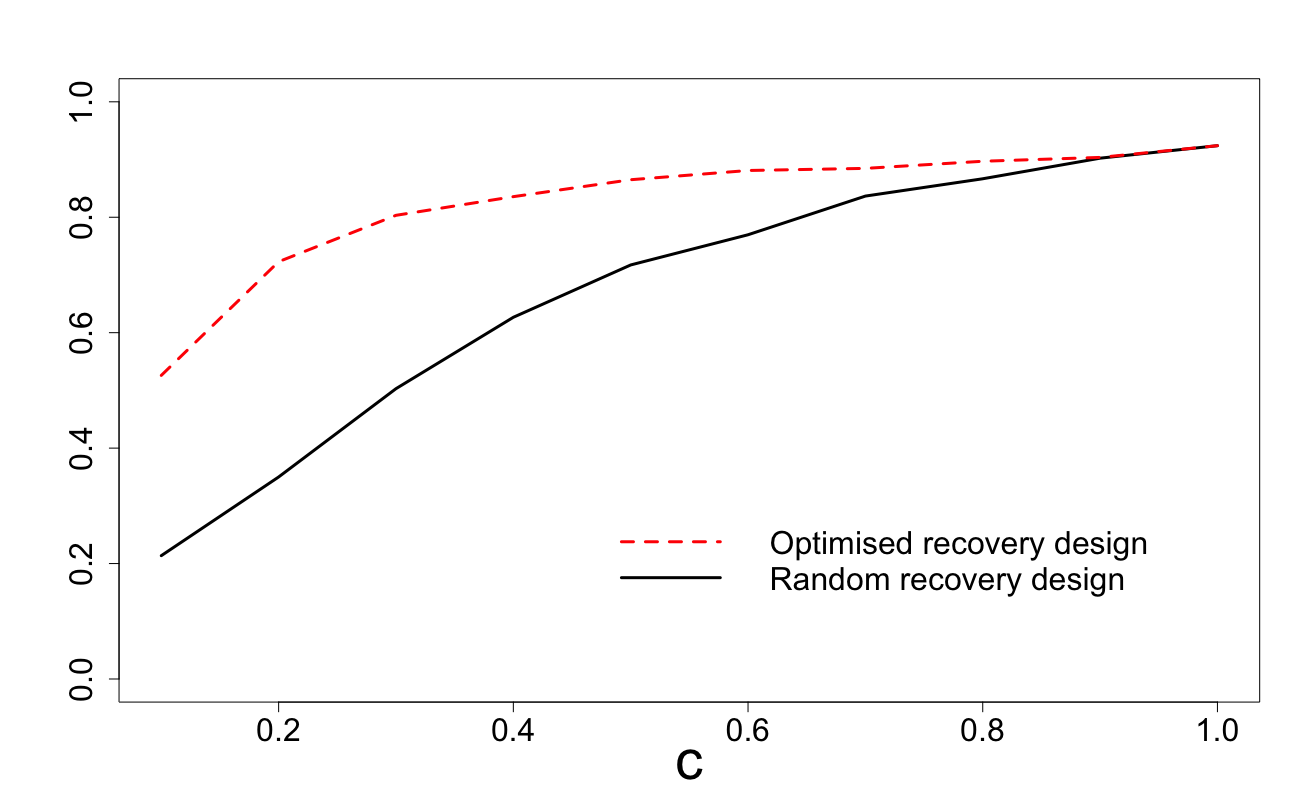}
  \centering
  \includegraphics[width=0.45\linewidth]{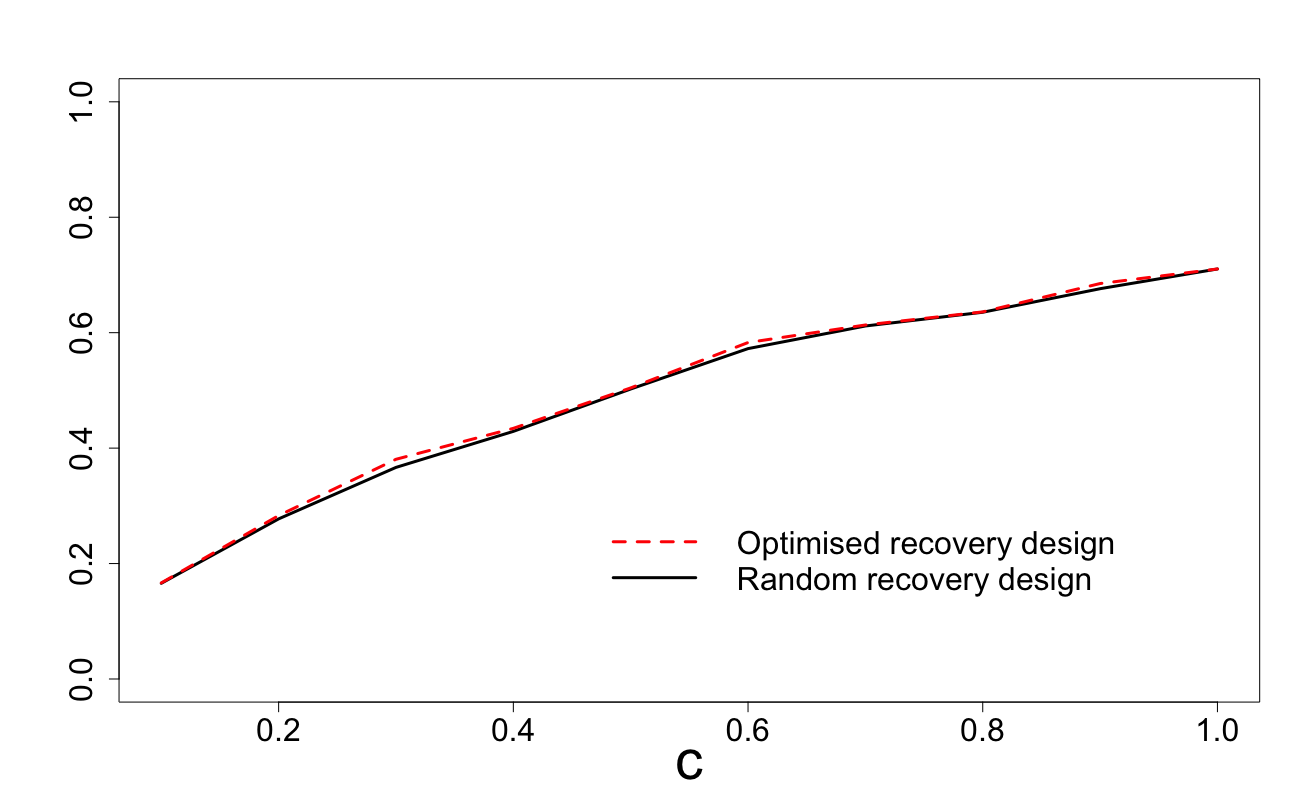}
\caption{Power for different recovery designs in Example 3 (L) and 4 (R).}
\label{fig7}
\end{figure}

{Again both mechanisms introduce approximately 30\% missingness into $y$.} In both examples, we apply test \eqref{SMF_test}, 
with $H_0: \bm{\psi}_A = 0$, over $10000$ replications of the data generation process. The results are depicted in Figure~\ref{fig7} (similar in style to Figure~\ref{fig6}), 
and we see for Example 3 the optimized recovery design produces significant benefits over the pure random recovery. However, for the MNAR mechanism in Example 4, the optimized recovery appears to produce identical results, in terms of power, to pure random recovery. {This highlights a case where the optimal recovery design is not qualitatively better than random recovery, but importantly does not perform worse.} 

\subsection{Assessing the robustness}\label{sec:robustness}

The optimality criteria we introduced in Sections~\ref{design1} and \ref{design2}, respectively, to find the optimal sampling regions $\cal{C}_A$, depend on the initial missing mechanism ${\rm Pr}(M=1 \,\vert \,\bm{X}=\bm{x},Y = y)$, the regression relation  ${\rm Pr}(Y\in dy \,\vert \, \bm{X}=\bm{x})$ and the distribution of $\bm{X}$. However, as $\bm{X}$ is fully observed, we can obtain good approximations for its distribution before designing the recovery sample.

Our robustness study is based on Example 1, with the true model having regression relation \eqref{regression_model_1}, i.e. 
$
Y|(X=x) \sim N(2-2x, 4),
$ 
and missing mechanism \eqref{first_logistic_example}, i.e. an expit model with linear predictor 
$ - 2 + 0.4 x - 0.15 y$. 

To assess the robustness of our proposed design methodology, we find optimal recovery regions where one parameter has been misspecified in turn. In particular, for each parameter we consider scenarios where we, respectively, add and subtract 10\% of its value, to see how the performance of the design is affected if the misspecification occurs in a neighborhood of the true values. To assess more severe misspecifications, we also consider scenarios where the parameters change signs and, for the coefficient of $y$ in the linear predictor of the missing mechanism, we also assess the effect of doubling the value to $-0.3$. 

Table 2 shows the power of each of these 16 misspecified optimal designs when applied to the true model, the power of the optimal design for the true parameter values, and the power of the random design, all based on 100,000 simulation replicates.

	\begin{table}[h!]
		\caption{Power for different misspecified optimal designs and recovery proportions $c_1=0.1,0.2,\ldots,1$. The true missing model has parameters $(-2,0.4,-0.15)$, and the true regression relation has parameters $(2,-2)$.}
		\centering
		\resizebox{15cm}{!}{\begin{tabular}{rlllllllllll}
			\hline
& Design & 0.1 & 0.2 & 0.3 & 0.4 & 0.5 & 0.6 & 0.7 & 0.8 & 0.9&1.0 \\ 
			\hline
& true optimal &0.287&0.463&0.595&0.672&0.724&0.758&0.767&0.796&0.809&0.830\\ 
& random &0.243&0.385&0.493&0.576&0.642&0.693&0.740&0.774&0.803&0.830\\ 
\hline
& Missing mechanism & & & & & & & & & & \\
&$(-2.2,0.4,-0.15)$ &0.280&0.459&0.584&0.668&0.721&0.757&0.765&0.791&0.809&0.830\\ 
& $(-1.8,0.4,-0.15)$ &0.287&0.464&0.568&0.654&0.680&0.711&0.744&0.781&0.803&0.830\\ 
& $(2,0.4,-0.15)$ &0.283&0.393&0.578&0.589&0.694&0.750&0.759&0.790&0.804&0.830\\ 
& $(-2,0.44,-0.15)$ &0.286&0.461&0.574&0.671&0.722&0.738&0.757&0.772&0.804&0.830\\ 
& $(-2,0.36,-0.15)$ &0.286&0.462&0.594&0.672&0.723&0.757&0.762&0.789&0.806&0.830\\ 
& $(-2,-0.4,-0.15)$ &0.252&0.390&0.495&0.591&0.670&0.720&0.757&0.777&0.804&0.830\\
& $(-2,0.4,-0.165)$ &0.287&0.460&0.584 &0.671&0.726&0.737&0.764&0.786&0.805&0.830\\
& $(-2,0.4,-0.135)$ &0.285&0.459&0.586 &0.672&0.723&0.738&0.766&0.781&0.803&0.830\\& $(-2,0.4,0.15)$ &0.252&0.391&0.501&0.597&0.665&0.719&0.755&0.781&0.807 &0.830\\ 
& $(-2,0.4,-0.3)$ &0.287&0.461&0.592&0.672&0.702&0.711&0.756&0.777&0.804&0.830\\
\hline
& Regression relation & & & & & & & & & & \\
& $(2.2,-2)$  &0.284&0.462&0.589&0.671&0.722&0.738&0.765&0.778&0.804&0.830\\ 
& $(1.8,-2)$  &0.285&0.463&0.594&0.648&0.703&0.730&0.755&0.783&0.805&0.830\\ 
& $(-2,-2)$  &0.285&0.462&0.581&0.671 &0.712&0.747&0.764&0.785&0.807&0.830\\ 
& $(2,-2.2)$  &0.287&0.463&0.575&0.629&0.703&0.709&0.745&0.776&0.802&0.830\\ 
& $(2,-1.8)$ &0.285&0.461&0.594&0.650&0.700&0.730&0.755&0.780&0.803&0.830\\ 
& $(2,2)$ &0.252&0.398&0.507&0.594&0.660&0.714&0.755&0.775&0.803&0.830\\ 
			\hline
		\end{tabular}}
	\end{table}

We can see from Table 2 that the misspecified optimal designs all do better than the random design, in particular when considering the more practically relevant scenarios of small values of the recovery proportion $c_1$. Many misspecified optimal designs show a similar performance to the true optimal design.

Looking closer at those misspecified optimal designs that perform not as well as the true optimal design, we find that these correspond to scenarios where the coefficient of $x$ or $y$ changed sign in either the missing mechanism or the regression relation, i.e. the parameter combinations $(-2,-0.4,-0.15)$ and $(-2,0.4,0.15)$ for the missing mechanism and $(2,2)$ for the regression relation. For example, when $c_1=0.1$, the power for these designs is only 0.252 compared with 0.287 for the true optimal design, but still higher than 0.243, the power of the random design.  

In conclusion, it seems that the optimal designs are highly robust to small misspecifications of the values of the unknown model parameters, to more severe changes such as the doubling of a value as long as the sign remains the same, and also to severe changes (including sign changes) in the intercepts $\alpha_0$ and $\beta_0$. 

The designs are less robust (but in this example still doing better than the random design) if the \textit{sign} of the coefficient of $x$ or $y$ is misspecified. When specifying the model for design search, it seems therefore most important to get the signs of these coefficients right, whereas the actual values are less crucial.

\section{Application to a real data example}

We also apply our methods to a scenario where our observed data values are not simulated from a known statistical distribution, using a study derived from the 1979 National Longitudinal Survey of Youth, commonly referred to as the NLSY79. This longitudinal survey, begun in 1979, interviewed a nationally representative sample of 12,686 young men and women in the US. 
From 1986, information on children born to women in the survey was also collected. For more information about the survey see \cite{mitra2011estimating}. We note here that our sole purpose is to establish potential gains in detecting the presence of MNAR in this setting where the data generating mechanism is unknown, and we do not seek to infer true causal effects. 

\begin{figure}[h]
\centering
  \includegraphics[width=0.45\linewidth]{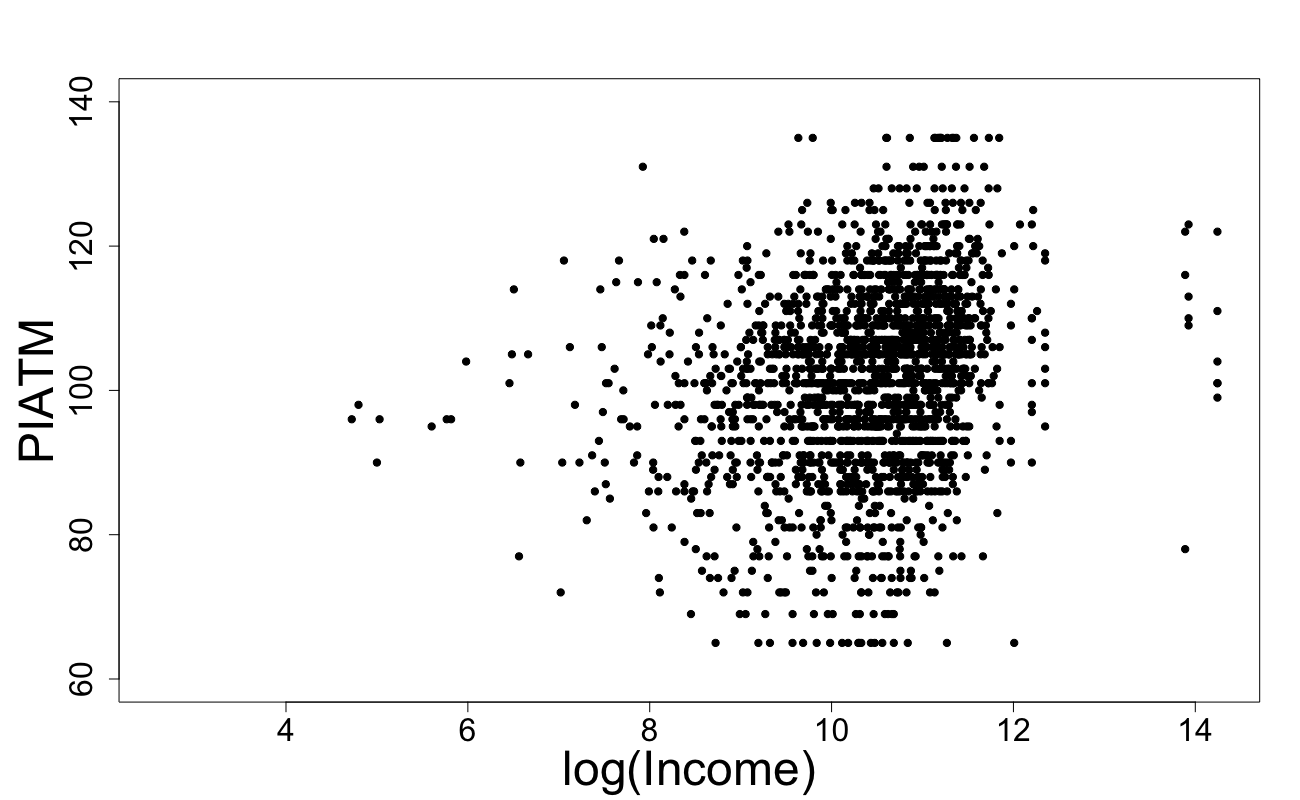}
  \includegraphics[width=0.45\linewidth]{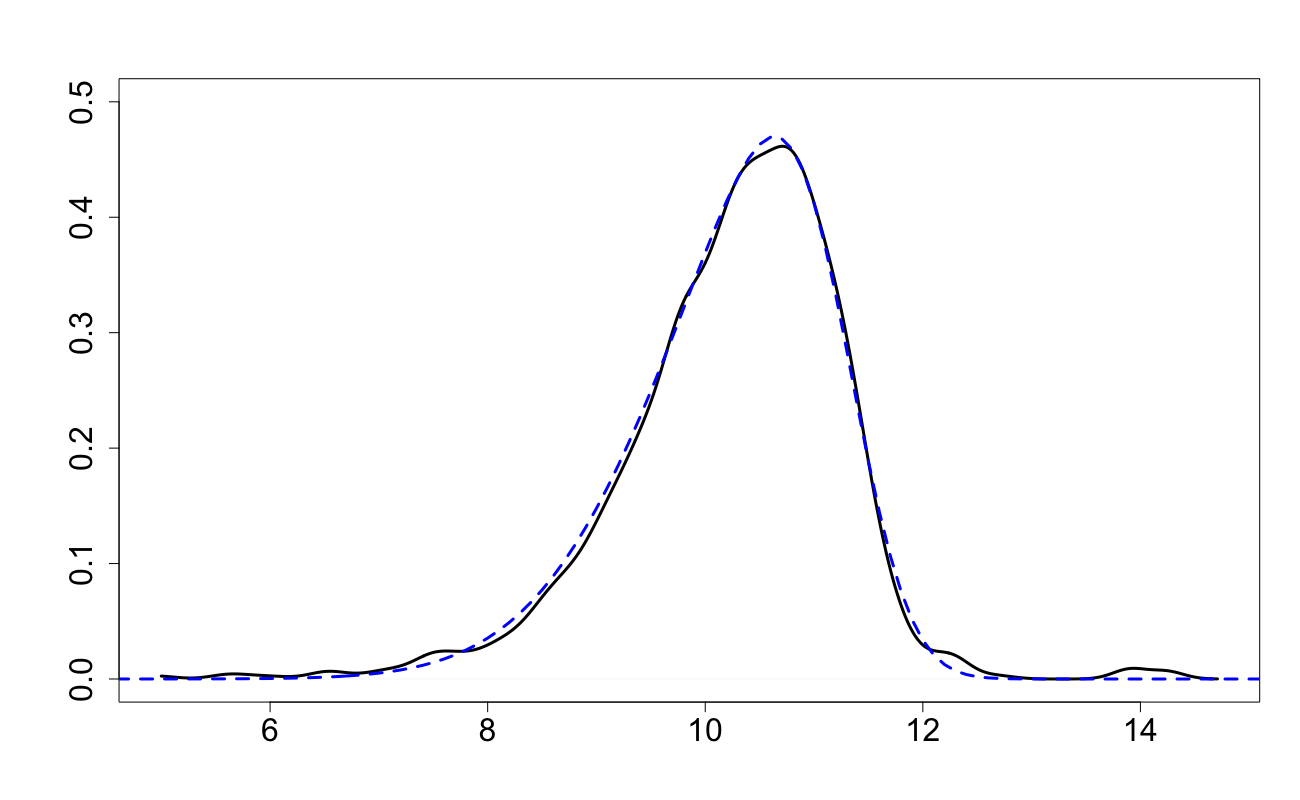}
    \caption{Peabody score vs log(income) (L); density of log(income) and skewed normal approximation (R)}
\label{PIATM and income}
\end{figure}

{Following \cite{mitra2011estimating} we subset on first born children only to avoid complications with family nesting. The resulting data set comprises 4888 observations.} {For our analysis model we consider a linear regression model of the form,
\bea
y_i = \beta_0 + \beta_1 x_i + \epsilon_i\,, \quad \epsilon_i \sim N(0,\sigma^2) \,,
\eea
where $y_i$ corresponds to the ith child's Peabody Maths score (PIATM) administered at 5 or 6 years of age (taken as a proxy for cognitive development), and $x_i$ corresponds to (the logarithm of) family income at birth, $i = 1, \ldots, 4888$. Since 
we assume covariates are fully observed, we focus on the subset of observations with observed family income, resulting in 3596 observations. We note alternative analysis models could be chosen but settled on this with evidence suggesting a relationship between these two variables \citep{cooper2021}.} Of the 3596 observations, PIATM is missing in 1640 cases (a 45.6\% missing rate). 
A scatterplot of the 1956 observed PIATM scores against the log of family income is depicted in Figure~\ref{PIATM and income} (L). The parameters that determine the conditional distribution of $y\vert x$ ($\beta_0,\beta_1,\sigma$) are unknown to us in this real data setting and will be estimated from the available data. Doing so results in the estimates $\beta_0\approx 69.06, \beta_1\approx 3.14  $ and $\sigma\approx13.14$. When designing our recovery regions, we will assume these are the true values. This seems reasonable following the analysis on robustness to misspecification of these parameters which is provided in Section~\ref{sec:robustness}. In Figure~\ref{PIATM and income} (R) we depict the density estimate of the log(income). A suitable candidate density is the skewed normal distribution with $\xi=11.3, \omega=1.4, \alpha=-3$ as indicated by the dashed blue line overlaid on the empirical density estimate.

\subsection{Simulation based on the complete case subsample}\label{subsec_6}

In this section, we subset our data on the complete case subsample, which comprise 1956 units. We introduce missingness back into the outcome, $y$, using the following mechanism: 
\be\label{real_data_model}
{\rm Pr}(M=1|{\rm PIATM}=y,{ \log({\rm income})} = { x})=\frac{\exp(\alpha_0+\alpha_1x +\alpha_2 y)}{1+\exp(\alpha_0+\alpha_1x +\alpha_2 y)} \,.
\ee
We specifically consider two scenarios for the missing data mechanism:
\begin{itemize}
\item[A)] $(\alpha_0, \alpha_1, \alpha_2) = (50,-5,0.016)^T$ and B) $(\alpha_0, \alpha_1, \alpha_2) = (-52,5,-0.017)^T$\,.
\end{itemize}
The parameter values $\alpha_i$ are chosen to introduce approximately 45\% missing data into the outcome (similar to the amount present in the original data). As before, we assume we are able to recover a proportion of the missing $y_i$ values.

We apply our methodology to optimizing the power of the test described in Section 2.2 above. To do this, we repeatedly generate new samples from the complete cases by resampling rows of the data with replacement, i.e. bootstrap the row indices. The number of resamples is set to 1956 (the number of complete cases). In each scenario, we then introduce missing values into the outcome using the mechanisms described above.

\begin{figure}[h]
\centering
  \includegraphics[width=0.45\linewidth]{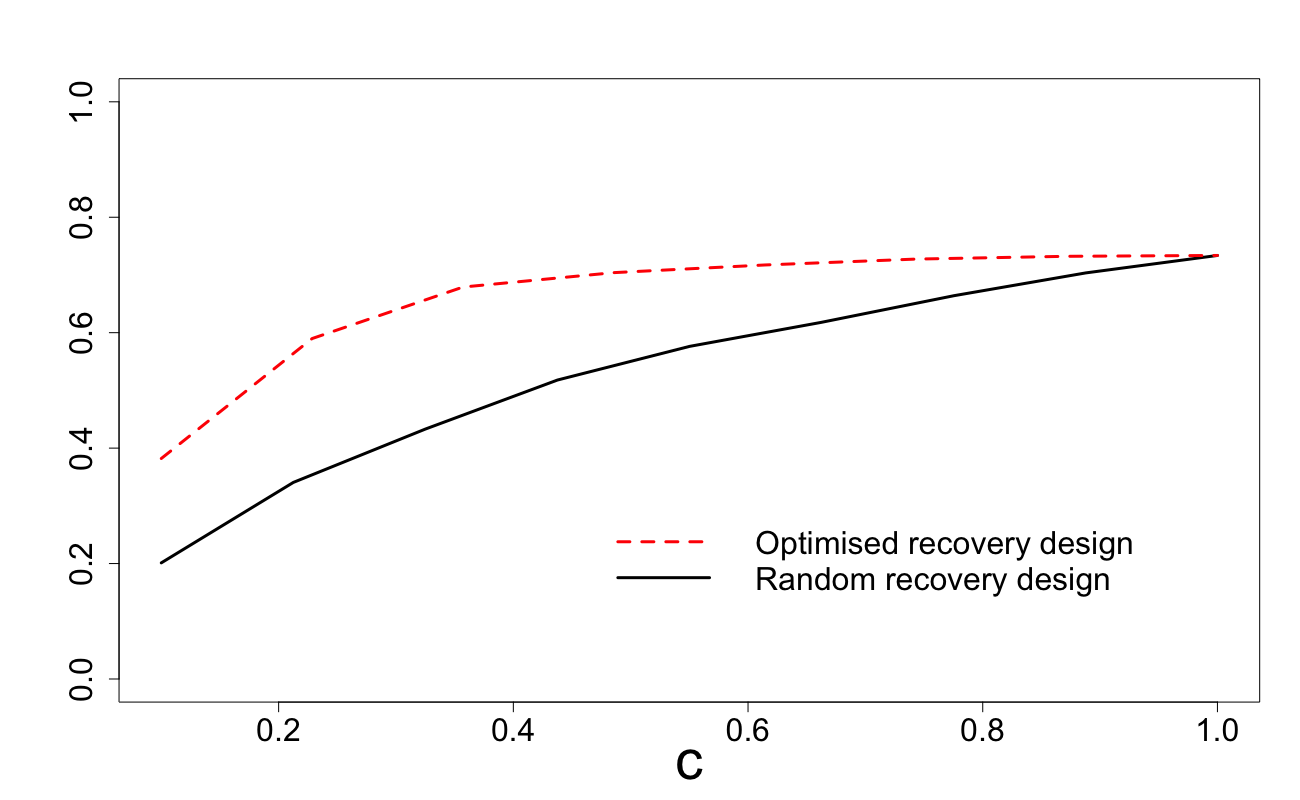}
  \includegraphics[width=0.45\linewidth]{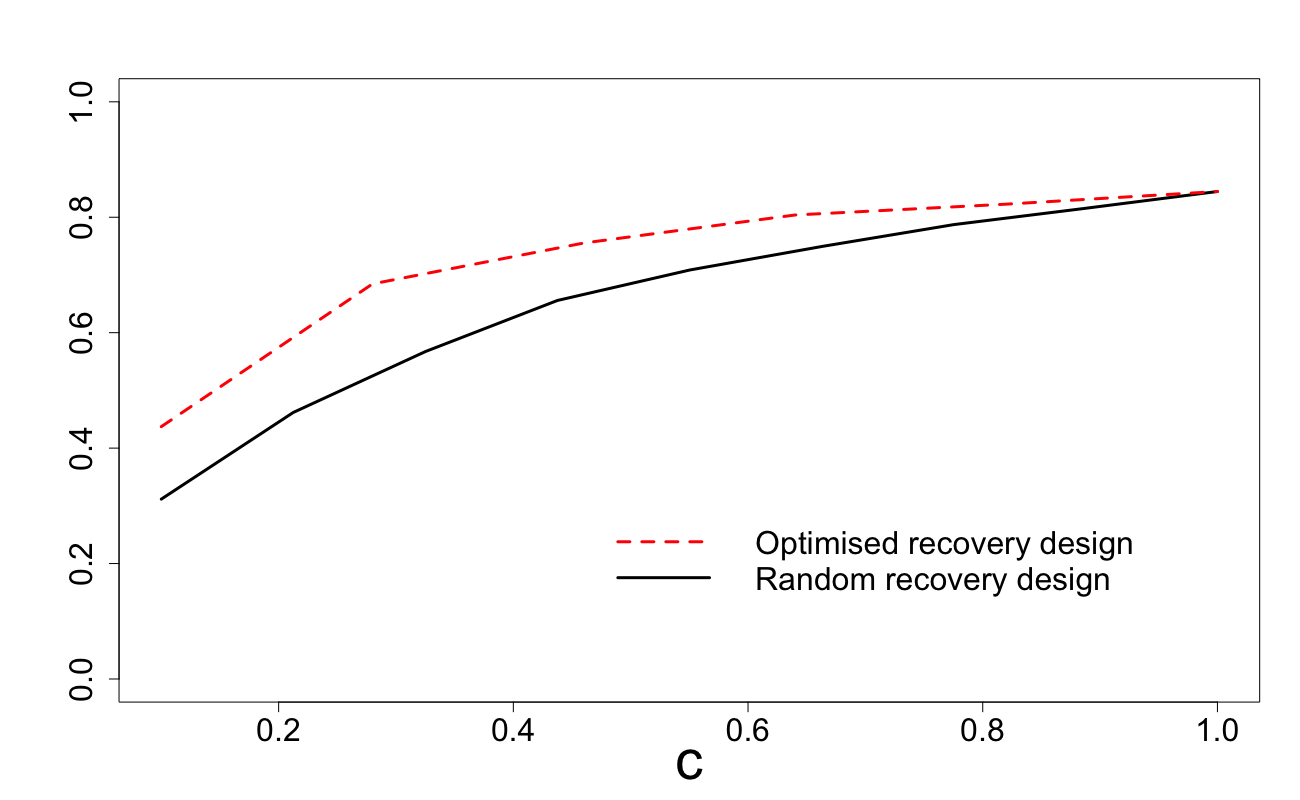}
    \caption{A comparison of power using the optimal recovery design versus a random recovery design in Scenario A (L)  and Scenario B (R).  }
\label{ccsimresults}
\end{figure}

Our results are presented in Figure \ref{ccsimresults}. 
We see a substantial increase in power with small recovery proportions. For example in Scenario A, for a recovery proportion $c_1 = 0.2$, random recovery has a power of $\approx 0.35$, compared to a power of 0.56 using the optimal recovery design. We also see for $c_1 = 0.4$ the power using the optimal recovery design is $\approx 0.7$, and to achieve a similar power using random recovery, a recovery proportion of almost 1 would be required, impractical in many settings. Scenario B shows similar gains for optimal recovery designs.

{In results not reported here, we compared power between the different recovery strategies when we assumed the distribution of log(income) was normal (a commonly used assumption). While this is a poorer approximation to this covariate's distribution, we saw little difference to the results reported in Figure \ref{ccsimresults}. This, combined with our robustness investigation in Section~\ref{sec:robustness}, gives us confidence that our gains would still be evident under modest model misspecification. 

\subsection{Application to the full data sample}

While we cannot evaluate potential gains for the full data sample, it would be interesting to see how different our recovery designs are from the standard random recovery design under different MNAR mechanisms. In order to postulate our MNAR mechanism we assume the model in \eqref{real_data_model} that takes parameter values in Scenario A and Scenario B of Section~\ref{subsec_6}. In Figure~\ref{Des2}, as a function of the recovery proportion $c=c_1$, we depict the upper and lower endpoints of the optimized recovery region ${\cal C}_{A}$ obtained from Algorithm 2b for Scenario A (dashed red line) and Scenario B (dotted blue line); note ${\cal C}_{A}$ is an interval $[a,b]$ since $p=1$. In this figure, the solid black lines correspond to the $0.1\%$ and $99.9\%$ quantiles of the covariate log(income) and can be useful in indicating how the design with optimized ${\cal C}_{A}$ differs from the standard random recovery design. We see that the optimized recovery designs for Scenario A and Scenario B are significantly different from the random recovery.

\begin{figure}[h]
\centering
  \includegraphics[width=0.6\linewidth]{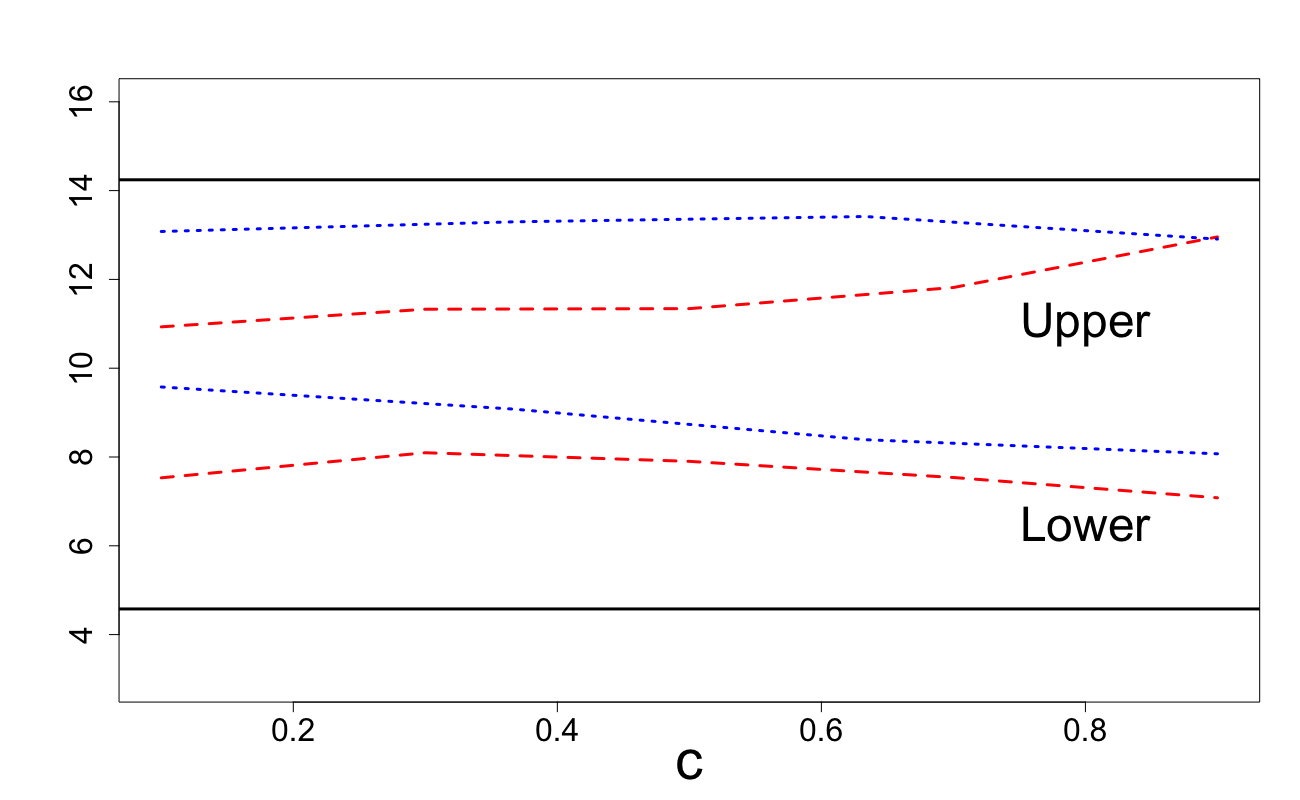}
  \caption{Endpoints of ${\cal C}_{A}$ for Scenario A (dashed red) and Scenario B (dotted blue).}
    \label{Des2}
\end{figure}


\section{Concluding remarks}

This research provides comprehensive insight into testing for the presence of MNAR utilizing a recovery sample, a little explored area. 
Firstly, 
our mathematically expressions for the missing data mechanism based on the subsample of observed plus recovered data permits 
principled inferences to be made in this setting. Notably, for the commonly used expit model, irrespective of the recovery design used, we determine that there is a shift in the intercept of the model by $\log{(c^{*})}$ while the other coefficients in the linear predictor, and indeed the form of the model, remain unchanged. {For other link functions, to preserve the original mechanism, the model must be fit to the recovered plus only a randomly sampled pre-specified proportion of the observed data.} Consequently, the SMF test \eqref{SMF_test} with $H_0:~\alpha_{p+1}=0$ based on the augmented data will reliably test for the presence of MNAR in the original data, giving analysts confidence in their results. Secondly, using experimental design methods to construct the recovery sample allows inferences to be optimized. 
In particular, 
constructing recovery designs that 
minimize the variance of the MLE for $\alpha_{p+1}$ ($D_1$-optimality) increases efficiency and thereby improves the power of the SMF test \eqref{SMF_test}. We note 
the equivalence to $T$-optimality \citep{atkinson1975design} here, 
and 
the potential to considerably outperform random recovery sampling merit their use.

Overall, this research provides a unified approach for detecting MNAR missingness in an incomplete data set in practice, combining a design strategy with a corresponding reliable, and well understood, test for MNAR. A well-chosen recovery design can achieve higher power to detect MNAR compared with random recovery sampling (for fixed recovery proportions). Similarly, if the power is fixed in advance, an efficient recovery design can result in a smaller proportion of missing responses needing to be followed up to achieve this power, thus reducing costs. 

In this work, we assume that a unit that is followed up will respond. 
However, there may be situations where this assumption is not realistic. We are planning to address this problem of potential partial recovery in future research to open up the methodology to further application areas, thus further increasing applicability and potential impact.  


\bibliography{paper-ref}

\section*{Appendix}
\subsection*{Proof of Lemma 1.}
We begin the proof by firstly assuming the condition in \eqref{key_condition} holds and $c_2=1$. For this case, we have ${M_A=M\vert \{{\cal M}_O\cup {\cal M}_R \}}$. Recognizing that events $\{ M = 0 \}$ and $\{ M = 1 \}$ are mutually exclusive, we can deduce by definition events ${\cal M}_O$ and ${\cal M}_R$ are also mutually exclusive. Thus, 
\bea
{\rm Pr}(M_A = 1) &= & {\rm Pr}(M = 1|\{{\cal M}_O\cup {\cal M}_R \}) \\
& = & \frac{{\rm Pr}(\{M = 1 \} \cap \{{\cal M}_O\cup {\cal M}_R \})}{{\rm Pr}(\{{\cal M}_O\cup {\cal M}_R \}})\\
& = & \frac{{\rm Pr}(\{M = 1 \} \cap \{{\cal M}_O\cup {\cal M}_R \})}{{\rm Pr}(\{M = 1 \} \cap \{{\cal M}_O\cup {\cal M}_R \}) + {\rm Pr}(\{M = 0 \} \cap \{{\cal M}_O\cup {\cal M}_R \}))}\\
& = & \frac{{\rm Pr}(\{M = 1 \} \cap {\cal M}_R)}{{\rm Pr}(\{M = 1 \} \cap {\cal M}_R) + {\rm Pr}(\{M = 0 \} \cap {\cal M}_O))} \\
& = & \dfrac{c_1\cdot{\rm Pr}(M=1)}{c_1\cdot {\rm Pr}(M=1)+{\rm Pr}({\cal M}_O )}\,.
\eea
If instead of condition \eqref{key_condition}, we assume that condition \eqref{key_condition2} holds and $c_2$ is not necessarily equal to one, then we need to modify the claim ${M_A=M\vert \{{\cal M}_O\cup {\cal M}_R \}}$ to account for the random subsample in the observed and recovered data. Let $U$ be a uniform random variable on $[0,1]$ and define the event ${\cal B}=\{U\leq c_1\cdot {\rm Pr}(M=1)/{\rm Pr}({\cal M}_R ) \} $. Then $M_A=M\vert \{ \{ {\cal M}_O\cap \{U\leq c_2\} \}\cup \{ {\cal M}_R \cap {\cal B}  \} \}$. The result then follows from almost identical calculations to the above.\\

\subsection*{Proof of Lemma~\ref{auxillary_lem1}.}
It follows for $\bm{x} \in  {\cal C}_{A}$:
\bea
{\rm Pr}(\bm{X}_R\in \bm{dx},Y_R \in dy) = {\rm Pr}(\bm{X}\in \bm{dx},Y \in dy\vert{\cal M}_R  )
&=& \frac{{\rm Pr}(\bm{X}\in \bm{dx},Y \in dy,{\cal M}_R  )}{ {\rm Pr}({\cal M}_R )}\\
&=& \frac{{\rm Pr}(\bm{X}\in \bm{dx},Y \in dy,M=1 )}{ {\rm Pr}({\cal M}_R )} \,.
\eea
The result then follows by applying rules of conditional probability in the numerator. The same approach can be used to derive the second result of the lemma for ${{\rm Pr}(\bm{X}_O\in \bm{dx},Y_O \in dy})$.

\subsection*{Proof of Theorem 1}
\subsubsection*{Auxiliary lemma}
The following auxiliary lemma will be used to prove Theorem~1.

\begin{lemma}\label{auxillary_lem2}
\begin{eqnarray*}
{\rm Pr}(M_A = 1|Y_A=y,\bm{X}_A = \bm{x} )={\rm Pr}(M_A=1) \times \frac{{\rm Pr}(\bm{X}_R \in \bm{dx}, Y_R  \in dy)}{{\rm Pr}(\bm{X}_A \in \bm{dx}, Y_A  \in dy)} \,.
\end{eqnarray*}
\end{lemma}


\subsubsection*{Proof of Theorem 1}
Combining the results of Lemma~\ref{auxillary_lem1} with Lemma~\ref{auxillary_lem2} provides
\begin{eqnarray*}
 {\rm Pr}(M_A = 1|\bm{X}_A =\bm{x}, Y_A = y)=  
\frac{{\rm Pr}(M_A=1){\rm Pr}(M=1|\bm{X} =\bm{x},Y=y){\rm Pr}(\bm{X} \in\bm{dx}, Y \in dy)}{A+ B} \,,
\end{eqnarray*}
where
\begin{eqnarray*}
A&=&{\rm Pr}(M_A=1){\rm Pr}(M=1|\bm{X} =\bm{x}, Y=y){\rm Pr}(\bm{X} \in\bm{dx}, Y \in dy) \\
B&=& \frac{{\rm Pr}({\cal M}_R )}{{\rm Pr}({\cal M}_O )}{\rm Pr}(M_A=0){\rm Pr}(M=0|\bm{X} =\bm{x}, Y=y) {\rm Pr}(\bm{X} \in \bm{dx}, Y\in dy) \,.
\end{eqnarray*}
 
 By cancelling the common term ${\rm Pr}(\bm{X} \in \bm{dx}, Y \in  dy)$ present in the numerator and denominator, and performing further rearrangement, we obtain:
\begin{eqnarray*}
 {\rm Pr}(M_A = 1|\bm{X}_A =\bm{x}, Y_A = y)= \frac{{\rm Pr}(M=1|\bm{X} =\bm{x},Y=y){\rm Pr}(M=0, \bm{X} \in {\bf {\cal C}}_{A} )}{C+D} \,,
\end{eqnarray*}
 where
 \begin{eqnarray*}
         C&=&{\rm Pr}(M=1|\bm{X} = \bm{x}, Y=y){\rm Pr}({\cal M}_O )\\
         D&=& \frac{{\rm Pr}(M_A=0)}{{\rm Pr}(M_A=1)} \cdot {\rm Pr}({\cal M}_R )\cdot {\rm Pr}(M=0|\bm{X} = \bm{x}, Y=y) \,.
 \end{eqnarray*}
From Lemma~1, one notices the following relation:
\begin{equation*}
\frac{{\rm Pr}(M_A=0)}{{\rm Pr}(M_A=1)} = \frac{c_2\cdot{\rm Pr}({\cal M}_O )}{c_1\cdot {\rm Pr}(M=1)} \,.
\end{equation*}
By then cancelling the common term ${\rm Pr}({\cal M}_O)$, and dividing the numerator and denominator by ${\rm Pr}({\cal M}_R)$, we obtain:
\begin{eqnarray*}
 {\rm Pr}(M_A = 1|\bm{X_A} = \bm{x}, Y_A = y)= 
 \frac{c^{*}{\rm Pr}(M=1|\bm{X} = \bm{x},Y=y)}{c^{*}{\rm Pr}(M=1|\bm{X} = \bm{x},Y=y)+ {\rm Pr}(M=0|\bm{X} = \bm{x}, Y=y)}
\end{eqnarray*}
where, 
\begin{equation*}
 c^{*}=\frac{c_1\cdot {\rm Pr}(M=1)}{c_2\cdot {\rm Pr}({\cal M}_R )} \,.
\end{equation*}

\end{document}